\documentclass[a4paper,aps,prd,showpacs,preprintnumbers,twocolumn,nofootinbib,superscriptaddress]{revtex4-1}

\usepackage[english]{babel}

\usepackage[utf8x]{inputenc}
\usepackage[T1]{fontenc}
\usepackage{ucs}
\usepackage{microtype}
\usepackage{newtxtext,newtxmath}
\usepackage{xcolor}

\usepackage{url}
\usepackage{hyperref}

\usepackage{xspace}
\usepackage{lastpage}
\usepackage{stmaryrd}

\usepackage{graphicx}
\usepackage{empheq}
\usepackage{ifthen}
\usepackage{tensor}
\usepackage{paralist}
\usepackage{mathrsfs}
\usepackage{extarrows}
\usepackage{verbatim}
\usepackage{upgreek}
\usepackage{wasysym}
\usepackage{lettrine}

\usepackage{natbib}
\usepackage[capitalise]{cleveref}
\crefname{section}{Sec.}{Secs.}
\Crefname{section}{Section}{Sections}
\usepackage{slashed}

\newcommand\ph{\ensuremath{\varphi}}

\newcommand\define{\equiv}

\newcommand\vect[1]{\boldsymbol{#1}}

\newcommand\ex[1]{\mathrm{e}^{#1}}
\renewcommand\i{\ensuremath{\mathrm{i}}}

\newcommand{\order}{\mathcal{O}}

\newcommand\e[1]{_{\mathrm{#1}}}
\newcommand\h[1]{^{\mathrm{#1}}}

\newcommand{\dd}{\mathrm{d}}

\newcommand{\ddf}[3][]{\frac{\dd^{#1} #2}{\dd {#3}^{#1}}}

\renewcommand\lim[2]{\underset{ #1 \rightarrow #2 }{ \mathrm{lim} } \,}

\newcommand{\delimiters}[4][]{
\ifthenelse{ \equal{#1}{1} }{  #2 #3 #4  }
					{ \ifthenelse{\equal{#1}{2}}{ \big#2 #3 \big#4 }
						{ \ifthenelse{\equal{#1}{3}}{ \Big#2 #3 \Big#4 }
							{ \ifthenelse{\equal{#1}{4}}{ \bigg#2 #3 \bigg#4 }
								{ \ifthenelse{\equal{#1}{5}}{ \Bigg#2 #3 \Bigg#4 }
									{ \left#2 #3 \right#4 }
								}
							}
						}
					}
													}

\newcommand{\pa}[2][]{\delimiters[#1]{(}{#2}{)}}
\newcommand{\pac}[2][]{\delimiters[#1]{[}{#2}{]}}

\newcommand{\ev}[2][]{\delimiters[#1]{\langle}{#2}{\rangle}}

\newcommand{\Lagrangian}{\mathcal{L}}
\newcommand{\EOM}{\mathcal{E}}
\newcommand{\propagator}{\mathcal{D}}
\newcommand{\bph}{\bar{\varphi}}
\newcommand{\dph}{\delta\varphi}
\newcommand{\hh}{\hat{h}}
\newcommand{\bg}{\bar{g}}
\newcommand{\bG}{\bar{G}}
\newcommand{\bX}{\bar{X}}
\newcommand{\massmatrix}{\mathsf{M}}
\newcommand{\amplitudematrix}{\mathsf{A}}
\newcommand{\kineticmatrix}{\mathsf{K}}
\newcommand{\diag}[1]{\slashed{#1}}
\newcommand{\tp}{\gamma}
\newcommand{\ta}{\Gamma}

\begin{document}

\title{Scalar and tensor gravitational waves}

\author{Charles Dalang}
\email{charles.dalang@unige.ch}
\affiliation{D\'{e}partement de Physique Th\'{e}orique, Universit\'{e} de Gen\`{e}ve,\\
24 quai Ernest-Ansermet, 1211 Gen\`{e}ve 4, Switzerland}

\author{Pierre Fleury}
\email{pierre.fleury@uam.es}
\affiliation{Instituto de F\'isica Te\'orica UAM-CSIC,
Universidad Aut\'onoma de Madrid,\\
Cantoblanco, 28049 Madrid, Spain}

\author{Lucas Lombriser}
\email{lucas.lombriser@unige.ch}
\affiliation{D\'{e}partement de Physique Th\'{e}orique, Universit\'{e} de Gen\`{e}ve,\\
24 quai Ernest-Ansermet, 1211 Gen\`{e}ve 4, Switzerland}

\begin{abstract}
In dark-energy models where a scalar field is nonminimally coupled to the spacetime geometry, gravitational waves are expected to be supplemented with a scalar mode. Such scalar waves may interact with the standard tensor waves, thereby affecting their observed amplitude and polarization. Understanding the role of scalar waves is thus essential in order to design reliable gravitational-wave probes of dark energy and gravity beyond general relativity.
In this article, we thoroughly investigate the propagation of scalar and tensor waves in the subset of Horndeski theories in which tensor waves propagate at the speed of light.
We work at linear order in scalar and metric perturbations, in the eikonal regime, and for arbitrary scalar and spacetime backgrounds. We diagonalize the system of equations of motion and identify the physical tensor mode, which differs from the metric perturbation.
We find that interactions between scalar and tensor waves generally depend on the scalar propagation speed. If the scalar waves are luminal or quasiluminal, then interactions are negligible. In the subluminal case, scalar-tensor interactions are effectively suppressed due to the incoherence of the wave's phases.
\end{abstract}

\date{\today}
\preprint{IFT-UAM/CSIC-20-136}
\maketitle

\section{Introduction}

\lettrine{A}{fter} more than a hundred years of general relativity (hereafter GR), deviations arising from its most popular alternatives remain elusive to observations~\cite{Will:2014kxa}. While it is always enough motivation to challenge the currently accepted theory, one should better know where deviations from appealing alternatives may appear. Promising candidates have emerged in the past two decades in light of cosmic acceleration~\cite{Perlmutter:1997zf, Riess:1998cb}, which is well described by GR with a cosmological constant but which is poorly understood from a theoretical point of view~\cite{Martin:2012bt}. 

Of these viable alternatives, Horndeski theories form a natural extension of GR, easy to study and featuring an extra scalar propagating degree of freedom, which will eventually be the main focus of this article. They form the most general four-dimensional Lorentz invariant set of scalar-tensor theories that lead to second-order equations of motion~\cite{Horndeski:1974wa}, thereby avoiding Ostrogradski instabilities~\cite{Ostrogradsky1850}. The idea of scalar fields that would mix with gravitational degrees of freedom naturally appears in the low-energy effective action of string theories or may emerge as a manifestation of Poincar\'e invariance in higher dimensions~\cite{deRham:2010eu, Jana:2020vov}. Alternatively, Horndeski theories may be motivated as a generic phenomenological attempt to model cosmic acceleration periods such as inflation~\cite{Kobayashi:2011nu} or dark energy~\cite{Kase:2018aps}. However, late-time deviations from GR are severely constrained by local tests of gravity such as from lunar laser ranging~\cite{Hofmann:2018myc} and Shapiro time delay~\cite{Bertotti:2003rm}. 

In that context, an attractive feature of Horndeski theories is their ability to recover GR locally through screening mechanisms. Well-known examples include the Vainshtein mechanism~\cite{Vainshtein:1972sx}, the chameleon scenario~\cite{2004PhRvD..69d4026K}, k-mouflage~\cite{Babichev:2009ee}, and see also Refs.~\cite{Brax:2019koq, Lombriser:2014ira} for more exotic scenarios.\footnote{See Ref.~\cite{Burrage:2020bxp} for recent mitigations concerning the effectiveness of Vainshtein screening in light of a UV completion.} In principle, this allows certain classes of Horndeski theories to escape local tests of gravity, but typically at the price of no longer explaining the cosmic acceleration directly through sufficiently large modified gravitational interactions~\cite{Wang:2012kj}.

\begin{figure}
    \centering
    \includegraphics[width=\columnwidth]{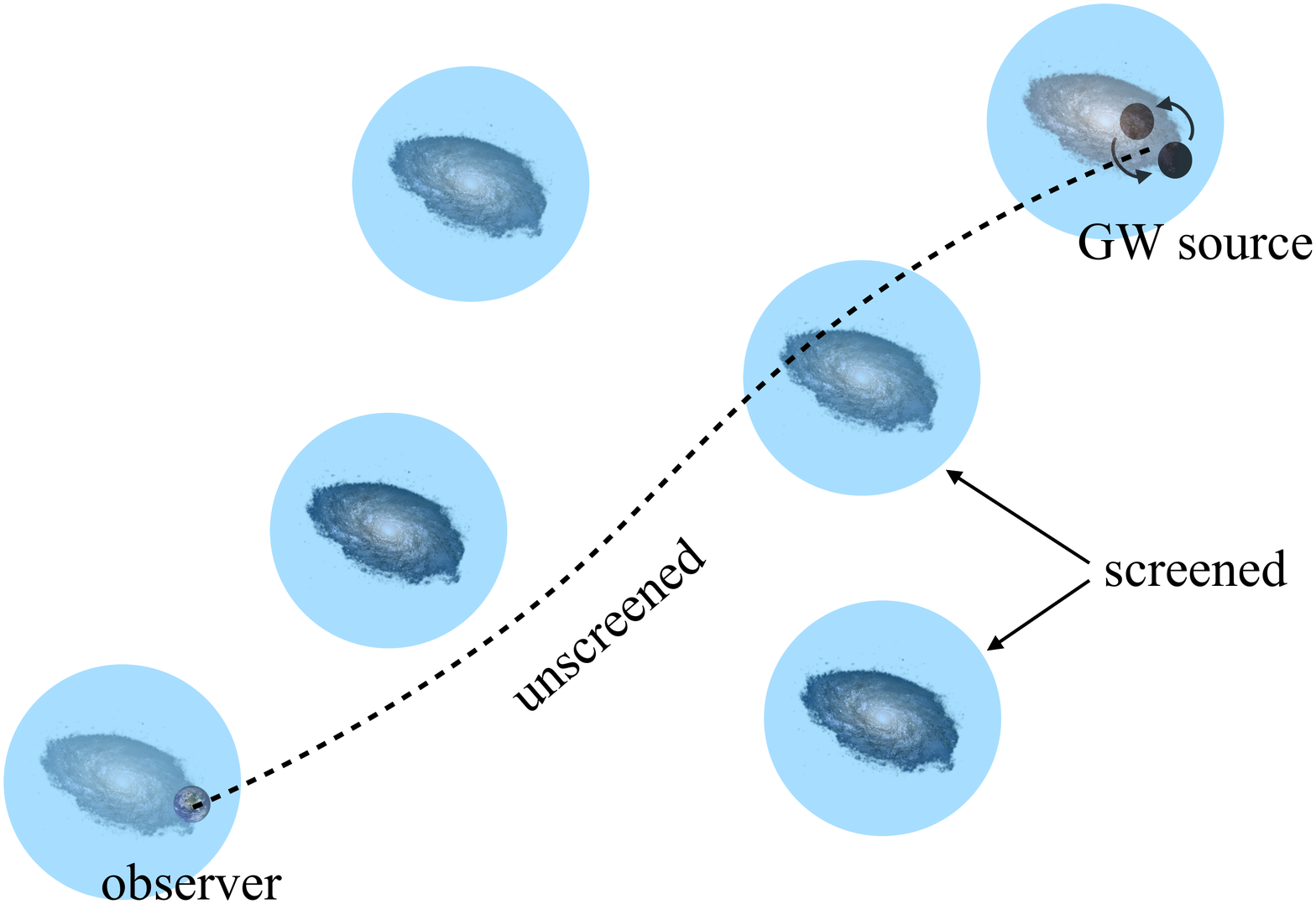}
    \caption{From its emission to its detection on Earth, a GW propagates through unscreened regions of the Universe, where it may interact with the scalar field that models dark energy. Note that this sketch is not to scale.}
    \label{fig:intergalacticmedium}
\end{figure}

If the deviations from GR are screened in regions where typical gravitation experiments are conducted, then the most promising area to look for them is the intergalactic medium. Gravitational waves (GWs) are a key candidate for that purpose, because they probe those inaccessible regions as they propagate through the Universe, as depicted in Fig.~\ref{fig:intergalacticmedium}.
The foremost example is the almost-simultaneous detection of GW170817~\cite{TheLIGOScientific:2017qsa} and GRB~170817A~\cite{Goldstein:2017mmi}, which has imposed stringent constraints on alternative theories of gravity. In practice, it has eliminated all theories that predict a deviation from luminal propagation of GWs~\cite{Lombriser:2015sxa, Ezquiaga:2017ekz, Creminelli:2017sry} for sources at reshift $z\lesssim 0.01$. In particular, it has posed severe challenges to a genuine explanation of cosmic acceleration from modified gravity~\cite{Lombriser:2016yzn}.\footnote{As a caveat, let us mention that this argument may not directly be applicable to perturbations describing the large-scale structure of the Universe as the energy scales involved in current GW experiments lie many orders of magnitude above cosmological scales~\cite{PhysRevD.98.023504, deRham:2018red}.}
Another consequence of the interaction between GWs and dark energy is the possible decay of the former into the latter, which practically rules out degenerate higher-order scalar-tensor theories (DHOST) as viable explanations of dark energy, and sets an upper limit on the kinetic-braiding parameter~\cite{Creminelli_2018, Creminelli:2019kjy}. 

Besides the GW speed and decay rate, constraints may also be obtained from the GW distance-redshift relation via the observation of standard sirens~\cite{1986Natur.323..310S, Holz:2005df}. While the gravitational and electromagnetic Hubble diagrams coincide in GR, they generally do not in alternative theories of gravity; in other words, $D\e{L}(z)\not= D\e{G}(z)$, where $D\e{L}, D\e{G}$ respectively denote the luminosity distance (measured with electromagnetic signals) and the gravitational distance (measured with standard sirens). Such a discrepancy has been envisaged as a promising probe of the cosmic evolution of the effective Planck mass~\cite{Lombriser:2015sxa, Amendola:2017ovw,PhysRevD.97.104066, Nishizawa:2017nef, Belgacem:2019pkk} and of the spatial clustering of dark energy~\cite{Garoffolo:2020vtd}.

The difference between $D\e{G}$ and $D\e{L}$ is usually interpreted in terms of an extra `friction' which non-GR fields would exert on GWs. However, as initially suspected in Ref.~\cite{Amendola:2017ovw, Dalang:2019fma} and demonstrated in Ref.~\cite{Dalang:2019rke}, in the most popular models the ratio $D\e{G}/D\e{L}$ only depends on \emph{local} properties of gravity at the emission and reception of the GW. Thus, the prospects of any program based on the difference between $D\e{G}$ and $D\e{L}$ should be quite limited by screening.\footnote{It should be noted that local experiments such as lunar laser ranging or Shapiro time delay do not constrain directly the local effective Planck mass, but rather the effective Newton constant appearing in the Poisson equation, or the gravitational slip. These couplings may differ in general~\cite{PhysRevLett.124.061101, Baker:2020apq}. However, for known screening mechanisms in Horndeski gravity such as the Vainshtein and chameleon scenarios, these couplings coincide in the deeply screened regime, leading to the absence of signature of the scalar field on the gravitational luminosity distance with respect to GR~\cite{Dalang:2019fma, Lagos_2020}.} This point has been mostly overlooked in the literature dedicated to standard-siren tests of gravity (see Ref.~\cite{Hogg:2020ktc} for a recent proposition to exploit it).

To be specific, Ref.~\cite{Dalang:2019rke} showed that, in Horndeski theories for which GWs propagate at the speed of light, $D\e{G}/D\e{L}=M(\ph\e{o})/M(\ph\e{s})$, where $M$ denotes the effective Planck mass and $\ph\e{s}, \ph\e{o}$ the dark-energy field at emission and observation of the GW. This turned out to be the only difference with GR. In particular, GWs still propagate along null geodesics of the background spacetime, and their polarization is parallel-transported.\footnote{Note that Ref.~\cite{Garoffolo:2019mna} reached a different conclusion on the polarization transport; See Appendix~E of Ref.~\cite{Dalang:2019rke} for an analysis of that discrepancy.}

Be that as it may, an important assumption in Ref.~\cite{Dalang:2019rke} was to neglect scalar waves. It is thus natural to wonder whether the aforementioned results -- distance formula, polarization transport -- hold when scalar waves are properly accounted for. Interactions between scalar and tensor waves are expected to arise in a realistic, inhomogeneous Universe which does not enjoy the exact symmetries of the idealized Friedmann-Lemaître-Robertson-Walker (FLRW) spacetime.\footnote{If scalar and tensor waves are treated as perturbations on an FLRW background, the standard scalar-vector-tensor decomposition~\cite{Peter:2013avv} implies that they involve independently at linear level; they do not at second order.} Scalar-tensor interactions may lead to energy transfers between the two sectors, thereby affecting, e.g., the observed $D\e{G}$. This question is particularly relevant if GWs are lensed, or propagate through nonlinear structures where screening may affect the scalar sector. Even if we could describe the background for GWs as an FLRW model with small perturbations, the accumulated impact of the latter over cosmic distances could still lead to significant effects. Any analysis of such setups that does not account for scalar waves is thus \textit{a priori} inconsistent.   
This article aims to complete this gap, by proposing a joint analysis of \emph{both} scalar and metric waves in Horndeski theories, within an arbitrary scalar and spacetime background.

The article is organized as follows. In \cref{sec:linearized_Horndeski}, after a brief introduction to the reduced Horndeski models, we linearize its equations of motion for both scalar and metric perturbations around an arbitrary background. In \cref{sec:scalar_tensor_waves}, we focus our analysis on wavelike perturbations, we derive the dispersion relations, and we identify the signatures of the three propagating degrees of freedom. In \cref{sec:Non_Interactions}, we show that the interactions between scalar and tensor GWs are negligible for luminal and quasiluminal scalar waves, defining a notion of scalar distance in the process. In the subluminal case, we discuss how scalar-tensor interactions may be suppressed due to the incoherence of the phases of the scalar and tensor waves. Finally, we summarize our results and conclude in \cref{sec:conclusion}.

We adopt the Misner-Thorne-Wheeler conventions~\cite{Misner:1974qy} for the metric signature and the Riemann tensor. Greek indices run from $0$ to $3$, Latin indices from $1$ to $3$. A comma indicates a partial derivative, $Z_{,\mu}\define\partial_\mu Z$, while a semicolon denotes a covariant derivative associated with the Levi-Civita connection, $Z_{\mu;\nu}\define \nabla_\nu Z_\mu$. Bold symbols represent Euclidean three-vectors; sans-serif symbols indicate matrices in the scalar-tensor field space. Symmetrization and anti-symmetrization of indices follow $Z_{(\mu\nu)}\equiv \frac{1}{2}(Z_{\mu\nu}+Z_{\nu\mu})$ and $Z_{[\mu\nu]}\equiv \frac{1}{2}(Z_{\mu\nu}-Z_{\nu\mu})$. A bar indicates a background quantity, while a hat indicates the trace-reversed counterpart of a rank-two tensor, $\hat{Z}_{\mu\nu}\define Z_{\mu\nu}-\frac{1}{2} (g^{\rho\sigma}Z_{\rho\sigma}) g_{\mu\nu}$. Units are such that $c=\hbar=1$.

\section{Linearized Horndeski models in the eikonal regime}\label{sec:linearized_Horndeski}

This section establishes the key equations governing linear perturbations of the gravitational field in Horndeski theories. The relevant action and equations of motion are given in \cref{subsec:Horndeski_theories}. \Cref{subsec:linearized_Horndeski} is dedicated to the linearization of the latter on an arbitrary background. In \cref{subsec:diagonalization} we identify the scalar and tensor modes by diagonalizing the kinetic term of the system of equations of motions.

\subsection{Reduced Horndeski theories}
\label{subsec:Horndeski_theories}

In this article, we focus on the subset of Horndeski theories in which metric perturbations propagate at the speed of light. This class of models will be referred to as \emph{reduced Horndeski theories} throughout the article. 

\subsubsection{Action}

Horndeski theories are an extension of GR featuring a scalar field $\ph$, which can interact nonminimally with the spacetime metric~$g_{\mu\nu}$. In the Jordan frame, Horndeski's action reads
\begin{align}\label{eq:Action}
S= S\e{m}[\psi\e{m},g_{\mu\nu}]+ S\e{g}[\varphi, g_{\mu\nu}] \, .
\end{align}
The matter sector~$S\e{m}$ depends on the matter fields $\psi\e{m}$, assumed to be minimally coupled to the spacetime metric $g_{\mu\nu}$.

In reduced Horndeski theories, the gravitational sector reads
\begin{align}
S\e{g}[\varphi,g_{\mu\nu}]
= \sum_{i=2}^4 S_i[\varphi,g_{\mu\nu}]
= \frac{M\e{P}^2}{2}\int_\mathcal{M} \dd^4x\sqrt{-g} \sum_{i=2}^4 \mathcal{L}_i\,,
\end{align}
where $M\e{P}\equiv 1/\sqrt{8\pi G}$ denotes the reduced Planck mass and the three Lagrangian densities~$\mathcal{L}_i$ are
\begin{align}
\label{eq:L_2}
\mathcal{L}_2 &= G_2(\varphi, X)\,,\\
\mathcal{L}_3 &= G_3(\varphi,X)\Box \varphi\,, \\
\label{eq:L_4}
\mathcal{L}_4 &= G_4(\varphi) R\,,
\end{align}
where $X=-\frac{1}{2}g^{\mu\nu} \varphi_{,\mu} \varphi_{,\nu}$ denotes the scalar field's kinetic term, $\Box\equiv g^{\mu\nu} \nabla_\mu \nabla_\nu$ the d'Alembertian operator, and $R$ the Ricci scalar. Note that conditions~\eqref{eq:L_2}-\eqref{eq:L_4} ensure that GWs propagate at light speed, as first formulated in Ref.~\cite{McManus:2016kxu}.

The functions $G_{2\dots 4}$ being mostly free, $S\e{g}$ encapsulates a wide class of scalar-tensor theories, notably GR itself, Jordan-Brans-Dicke theories~\cite{Brans:1961sx}, $f(R)$~\cite{DeFelice:2010aj}, quintessence~\cite{Tsujikawa:2013fta}, k-essence~\cite{ArmendarizPicon:2000dh, ArmendarizPicon:2000ah} and the cubic covariant galileon~\cite{Deffayet:2009wt}. In principle, this leaves enough freedom to locally screen deviations from GR, e.g. to evade Solar-System constraints.

\subsubsection{Equations of motion}
\label{subsubsec:EOM}

Requiring the action~\eqref{eq:Action} to be stationary with respect to variations of the scalar field~$\ph$, and variations of the inverse metric~$g^{\mu\nu}$, yields the equations of motion
\begin{align}
\label{eq:EoM_scalar}
\EOM_\ph &= 0 \ , \\
\label{eq:EoM_tensor}
\EOM_{\mu\nu} &= M\e{P}^{-2} T_{\mu\nu} \ ,
\end{align}
where
\begin{align}
\EOM_\ph
&\equiv
\frac{2M\e{P}^{-2}}{\sqrt{-g}}
\frac{\delta S\e{g}}{\delta\ph} \ ,\\
\EOM_{\mu\nu}
&\equiv
\frac{2M\e{P}^{-2}}{\sqrt{-g}}
\frac{\delta S\e{g}}{\delta g^{\mu\nu}} \ ,\\
T_{\mu\nu}
&\equiv -\frac{2}{\sqrt{-g}}
\frac{\delta S\e{m}}{\delta g^{\mu\nu}} \ .
\end{align}
The scalar~$\EOM_\ph$ and tensor $\EOM_{\mu\nu}$ may be split into three pieces each, following the division of $S\e{g}$ into $S_2, S_3, S_4$:
\begin{align}
\EOM_\ph
&= \sum_{i=2}^4 \EOM_\ph^{(i)} \ ,
\qquad
\EOM_\ph^{(i)}
\equiv \frac{2M\e{P}^{-2}}{\sqrt{-g}}
        \frac{\delta S_i}{\delta\ph} \ ;
\\
\EOM_{\mu\nu}
&= \sum_{i=2}^4 \EOM_{\mu\nu}^{(i)} \ ,
\qquad
\EOM_{\mu\nu}^{(i)}
\equiv \frac{2M\e{P}^{-2}}{\sqrt{-g}}
        \frac{\delta S_i}{\delta g^{\mu\nu}} \ .
\end{align}
The expressions of the scalar terms are
\begin{align}
\label{eq:E2_scalar}
\EOM_{\ph}^{(2)}
&= G_{2,\ph} 
    + G_{2,X} \Box\ph
    - 2X G_{2,X\ph}
    + G_{2,XX} \ph^{,\mu} X_{,\mu}
\\
\label{eq:E3_scalar}
\EOM_{\ph}^{(3)}
&= \left(
        2 G_{3,\ph} - 2X G_{3,\ph X}
        + G_{3,XX} \ph^{,\mu} X_{,\mu}
        + G_{3,X} \Box\ph
    \right) \Box\ph \nonumber\\
&\quad
    - 2X G_{3,\ph\ph}
    + 2G_{3,\ph X} \ph^{,\mu} X_{,\mu}
    + G_{3,XX} X_{,\mu} X^{,\mu} \nonumber\\
&\quad - G_{3,X} (\ph^{;\mu\nu} \ph_{;\mu\nu}
        + R^{\mu\nu} \ph_{,\mu} \ph_{,\nu})
\\
\label{eq:E4_scalar}
\EOM_{\ph}^{(4)}
&= G_{4,\ph} R \ ,
\end{align}
and those of the tensor terms are
\begin{align}
\label{eq:E2_tensor}
\EOM^{(2)}_{\mu\nu} &= -\frac{1}{2}
                    \pa{
                        G_{2,X} \ph_{,\mu} \ph_{,\nu}
                        + G_2 g_{\mu\nu}
                        } \\
\label{eq:E3_tensor}
\EOM^{(3)}_{\mu\nu} &= - G_{3,\ph} 
                    (\ph_{,\mu} \ph_{,\nu} + X g_{\mu\nu})
                    \nonumber\\
                    & - G_{3,X}
                        \pac{
                            \ph_{,(\mu} X_{,\nu)}
                            + \frac{1}{2} \Box\ph \,
                                \ph_{,\mu} \ph_{,\nu}
                            - \frac{1}{2}\ph^{,\rho} X_{,\rho}     g_{\mu\nu}
                            } \\
\label{eq:E4_tensor}
\EOM^{(4)}_{\mu\nu} &= G_4 E_{\mu\nu}
                    - G_{4;\mu\nu}
                    + \Box G_4 \, g_{\mu\nu} \,,
\end{align}
where $E_{\mu\nu}$ denotes the Einstein tensor. From \cref{eq:EoM_scalar}, it appears that the scalar equation of motion does not directly receive contributions from $S\e{m}$. This happens because matter is not directly coupled to $\ph$ in the Jordan frame. However, since $\EOM_\ph^{(3)}, \EOM_\ph^{(4)}$ feature the Ricci tensor, which is itself sourced by matter via \cref{eq:EoM_tensor}, the scalar field is actually sourced by matter just like the metric. This indirect coupling can be made explicit as follows. First, use \cref{eq:EoM_tensor} to express $R_{\mu\nu}$ and $R$ as
\begin{align}
R &= G_4^{-1}
    \pa{
        3\Box G_4
        + \EOM^{(2)}
        + \EOM^{(3)}
        - M\e{P}^{-2} T
        } 
\\
R_{\mu\nu} &= G_4^{-1}
            \pac{
                \pa{G_{4;\mu\nu}
                    + \frac{1}{2}\Box G_4 g_{\mu\nu}}
                - \hat{\EOM}^{(2)}_{\mu\nu}
                - \hat{\EOM}^{(3)}_{\mu\nu} 
                + M\e{P}^{-2} \hat{T}_{\mu\nu}
                } ,
\end{align}
with $\EOM^{(i)}\equiv g^{\mu\nu}\EOM^{(i)}_{\mu\nu}$, $T\equiv g^{\mu\nu} T_{\mu\nu}$, and where a hat indicates the trace-reversed counterpart of a tensor, for instance
\begin{equation}
\hat{\EOM}_{\mu\nu}^{(2)}
\equiv \EOM_{\mu\nu}^{(2)} 
    - \frac{1}{2} \EOM^{(2)} g_{\mu\nu} \ .
\end{equation}
Second, substitute the above in the expressions~\eqref{eq:E3_scalar}, \eqref{eq:E4_scalar} of $\EOM_\ph^{(3)}$ and $\EOM_\ph^{(4)}$.

Importantly, once the above operations are performed, the scalar equation of motion does not contain second-order derivatives of $g_{\mu\nu}$ any longer -- it expresses the dynamics of the scalar field \emph{only}. This does not happen with the tensor equation of motion~\eqref{eq:EoM_tensor}, because $\EOM_{\mu\nu}$ contains several terms with second derivatives of the scalar field. We shall come back to this issue at the level of linear perturbations in \cref{subsec:diagonalization}.

\subsection{Linear perturbations on an arbitrary background}
\label{subsec:linearized_Horndeski}

Consider small perturbations over an arbitrary background for both the scalar field and the metric,
\begin{align}
g_{\mu\nu}
&= \bg_{\mu\nu} + h_{\mu\nu}\,, \qquad ||h_{\mu\nu}||\ll 1 \ , \\
\varphi &= \bph + \dph \,, \qquad |\dph|\ll 1\ ,
\end{align}
where $||\ldots||$ can be any reasonable notion of norm. The goal of this subsection is to expand the equations of motion~\eqref{eq:EoM_scalar}, \eqref{eq:EoM_tensor} at first order in both $h_{\mu\nu}, \dph$. In Ref.~\cite{Dalang:2019rke}, we had neglected scalar perturbations~$\dph$ for simplicity, and considered metric perturbations only. We shall not make this approximation here, because we are precisely interested in how scalar and tensor perturbations may interact.

By definition, $\bar{\ph}, \bar{g}_{\mu\nu}$ are solutions of the equations of motion for some background energy-momentum distribution~$\bar{T}_{\mu\nu}$.
Thus, at first order in $\dph, h_{\mu\nu}$, we have
\begin{align}
\delta\EOM_{\ph} &= 0 \ , \\
\delta\EOM_{\mu\nu} &= M\e{P}^{-2}\delta T_{\mu\nu} \ .
\end{align}
From now on, \emph{we shall neglect GW sources},\footnote{The energy-momentum perturbation $\delta T_{\mu\nu}$ actually has two distinct contributions. On the one hand, any obvious addition to $T_{\mu\nu}$, say a black hole binary or a cosmic string, would be a physical source of GWs. We shall not consider such contributions because we want to focus on the \emph{propagation} of scalar and tensor waves; in particular, we assume that they are not externally sourced as they propagate. On the other hand, as a GW propagate through a Universe filled with matter fields, it generates a contribution to $\delta T_{\mu\nu}$ by perturbing the metric in $\bar{T}_{\mu\nu}$. For standard forms of cosmological fluids, that contribution to $\delta T_{\mu\nu}$ is a negligible $\order(\omega^0)$ term. Exceptions are relativistic fluids, for which that correction is $\order(\omega^1)$, but which have negligible impact on the propagation of GWs during matter or dark-energy dominated scenarios~\cite{Weinberg:2003ur}.}
i.e., assume $\delta T_{\mu\nu}=0$. The equations of motion then simply become $\delta\EOM_\ph=\delta\EOM_{\mu\nu}=0$.

The quantities $\delta\EOM_{\ph}, \delta\EOM_{\mu\nu}$ depend on both $\dph$ and $h_{\mu\nu}$, but also on their first and second derivatives. The many terms constituting $\delta\EOM_{\ph}, \delta\EOM_{\mu\nu}$ may thus be organized into groups, depending on the number of derivatives acting on $\dph, h_{\mu\nu}$. This is because we will eventually focus on rapidly oscillating fields, in which case more derivatives means a larger term. An abstract but compact way of writing the outcome of this classification is
\begin{equation}
\label{eq:linearized_EoMs_formal}
\begin{pmatrix}
\delta\EOM_\ph\\
\delta\EOM_{\mu\nu}
\end{pmatrix}
=
\pa{
    \kineticmatrix^{\alpha\beta}
        \bar{\nabla}_\alpha\bar{\nabla}_\beta
    + \amplitudematrix^{\alpha}
        \bar{\nabla}_\alpha
    + \massmatrix
    }
\begin{pmatrix}
\dph\\
\hh_{\rho\sigma}
\end{pmatrix} 
=
\begin{pmatrix}
0\\
0
\end{pmatrix}  ,
\end{equation}
where sans-serif symbols indicate matrices in the field superspace spanned by $\dph, \hat{h}_{\mu\nu}$. In practice, each symbol $\kineticmatrix^{\alpha\beta}, \amplitudematrix^{\alpha}, \massmatrix$ may be seen as an $11\times 11$ matrix.\footnote{Why $11$? Because the set $(\dph, \hh_{\mu\nu})$ has $11$ independent components: one for $\dph$, and $10$ for the independent components of the symmetric tensor~$\hh_{\mu\nu}$. This counting does not account for the gauge freedom.} Their respective role and interpretation should be clearer under the form of block matrices, as shown hereafter. Note that, in \cref{eq:linearized_EoMs_formal}, we chose to express the equations of motion in terms of the trace-reversed metric perturbation~$\hh_{\mu\nu}$; this choice leads to slightly simpler expressions for $\kineticmatrix^{\alpha\beta}, \amplitudematrix^{\alpha}$.

The \emph{kinetic} matrix
\begin{equation}
\label{eq:kinetic_matrix}
\kineticmatrix^{\alpha\beta}
=
\begin{pmatrix}
K\indices{_\ph^\ph^\alpha^\beta} &
K\indices{_\ph^\rho^\sigma^\alpha^\beta} \\
K\indices{_\mu_\nu^\ph^\alpha^\beta} &
K\indices{_\mu_\nu^\rho^\sigma^\alpha^\beta}
\end{pmatrix} ,
\end{equation}
whose explicit expressions of the blocks are given in \cref{subsubsec:kinetic_matrix_nondiagonal}, governs the second derivatives of the scalar and metric perturbations. It is therefore the core of the dynamical properties of the equations of motion. For wavelike perturbations, $\kineticmatrix^{\alpha\beta}$ governs the \emph{dispersion relations}. In particular, the diagonal components $K\indices{_\ph^\ph^\alpha^\beta}$ (resp. $K\indices{_\mu_\nu^\rho^\sigma^\alpha^\beta})$ would control the dispersion relation of scalar (resp. tensor) waves in the absence of tensor (resp. scalar) waves. The non-diagonal components encode kinetic mixing, i.e., how second derivatives of $\dph$ contaminate the equation of motion of $\hh_{\mu\nu}$ and vice-versa. Since these off-diagonal components are generally nonzero in reduced Horndeski theories, we conclude that $\dph, \hh_{\mu\nu}$ are not their actual degrees of freedom. We shall explicitly address this issue in \cref{subsec:diagonalization}.

The \emph{amplitude} matrix
\begin{equation}
\label{eq:amplitude_matrix}
\amplitudematrix^{\alpha}
=
\begin{pmatrix}
A\indices{_\ph^\ph^\alpha} &
A\indices{_\ph^\rho^\sigma^\alpha} \\
A\indices{_\mu_\nu^\ph^\alpha} &
A\indices{_\mu_\nu^\rho^\sigma^\alpha}
\end{pmatrix}
\end{equation}
rules the first derivatives of $\dph, \hh_{\mu\nu}$. For wavelike perturbations, $\amplitudematrix^{\alpha}$ controls the evolution of their amplitudes. The expressions of the blocks are given in \cref{subsubsec:amplitude_matrix_nondiagonal}. The diagonal terms are self-damping or self-amplification terms, while the off-diagonal terms encode the interactions between scalar and tensor waves. In other words, the latter tell us how energy is exchanged between scalar and tensor waves as they propagate.

Finally, the \emph{mass} matrix
\begin{equation}
\label{eq:mass_matrix}
\massmatrix
=
\begin{pmatrix}
M\indices{_\ph^\ph} &
M\indices{_\ph^\rho^\sigma} \\
M\indices{_\mu_\nu^\ph} &
M\indices{_\mu_\nu^\rho^\sigma}
\end{pmatrix}
\end{equation}
contains the terms with no derivatives. Such terms would be involved in the dispersion relation of scalar and tensor waves at next-to-next-to-leading order. We shall justify in \cref{subsec:wave_ansatze} that they can be neglected in the eikonal regime. Anticipating on this simplification, \emph{we neglect all masslike terms from now on}; thus, there is not need to explicitly compute their expressions in the scope of this article.

\begin{description}
\item[Example] Let us illustrate the above in the simple case of GR with minimally coupled quintessence, i.e.
$
\Lagrangian_2 = X ,
\Lagrangian_3 = 0 ,
\Lagrangian_4 = R .
$
In that case, omitting masslike terms,
\begin{align}
\delta\EOM_\ph
&=
\Box\dph - \bph_{,\mu}\hh\indices{^\mu^\nu_;_\nu}
\\
\delta\EOM_{\mu\nu}
&=
\frac{1}{2} \pac{
                2\hh\indices{_\rho_(_\mu^;^\rho_\nu_)}
                - \Box\hh_{\mu\nu}
                - \hh\indices{^\rho^\sigma_;_\rho_\sigma}
                    \bg_{\mu\nu}
                } \ ,
\end{align}
and hence
\begin{align}
\kineticmatrix^{\alpha\beta}
&=
\begin{pmatrix}
\bg^{\alpha\beta} & 0 \\
0 &
K\indices{_\mu_\nu^\rho^\sigma^\alpha^\beta}
\end{pmatrix} ,
\\
\amplitudematrix^{\alpha}
&=
\begin{pmatrix}
0 & -\bph^{,\rho}\bg^{\sigma\alpha} \\
0 & 0
\end{pmatrix} ,
\end{align}
with
\begin{align}
K\indices{_\mu_\nu^\rho^\sigma^\alpha^\beta}
&=
\frac{1}{2}\bigg[
                2\bg^{\alpha(\rho}\delta^{\sigma)}_{(\mu}\delta^\beta_{\nu)}
                - \delta^{(\rho}_\mu\delta^{\sigma)}_\nu\bg^{\alpha\beta}\nonumber\\&\qquad
                - \bg^{\alpha(\rho}\bg^{\sigma)\beta}\bg_{\mu\nu}
            \bigg] .
\end{align}
In this example, the kinetic matrix is diagonal. This reflects the minimal coupling between the scalar and tensor fields, which are the true degrees of freedom of GR with quintessence.
\end{description}

\subsection{Kinetic diagonalization and gauge fixing}
\label{subsec:diagonalization}

Unlike the above simple example, $\dph, \hat{h}_{\mu\nu}$ are generally not the actual propagating degrees of freedom of linearized Horndeski models. Finding those requires to diagonalize the kinetic term of \cref{eq:linearized_EoMs_formal}, which is the goal of this subsection. This can be done in two successive steps, which we sketch in \cref{subsubsec:kinetically_scalar_EOM,subsubsec:kinetically_tensor_EOM}. Further details on the actual operations can be found in \cref{subsubsec:diagonalization_operations}. Additional simplifications of the resulting equations of motion are obtained by imposing an analog of the harmonic gauge, as shown in \cref{subsubsec:harmonic}.

\subsubsection{Eliminating second derivatives of \texorpdfstring{$h_{\mu\nu}$}{hmunu} from \texorpdfstring{$\delta\EOM_\ph$}{delta E phi}}
\label{subsubsec:kinetically_scalar_EOM}

The first step of the diagonalization procedure consists in isolating an equation of motion for $\dph$ that does not contain any second derivatives of the metric perturbation. In other words, we aim here to remove the off-diagonal block~$K\indices{_\ph^\rho^\sigma^\alpha^\beta}$.

In fact, as already mentioned at the end of \cref{subsubsec:EOM}, this can even be achieved nonperturbatively by substituting the Ricci terms in $\EOM_\ph$ by their expression obtained from $\EOM_{\mu\nu}$. At the linear-perturbation level, that substitution is equivalent to combining $\delta\EOM_\ph, \delta\EOM_{\mu\nu}$ as follows,
\begin{align}
\label{eq:transformation_scalar_row}
\delta\EOM_\ph &\mapsto \delta\EOM_\ph + C^{\mu\nu}\delta\EOM_{\mu\nu} \ , \\
\label{eq:transformation_tensor_row}
\delta\EOM_{\mu\nu} &\mapsto \delta\EOM_{\mu\nu} \ ,
\end{align}
where we have introduced the tensor
\begin{equation}
\label{eq:definition_C}
C_{\mu\nu}
\define
\bG_{4}^{-1}
    \pac{
        \bG_{3,X} \pa{
                    \bph_{,\mu}\bph_{,\nu}
                    + \bX \bg_{\mu\nu}
                    }
        + \bG_{4,\ph} \bg_{\mu\nu}
        } .
\end{equation}
Thanks to the nontrivial identity (see \cref{subsubsec:diagonalization_operations})
\begin{equation}
K\indices{_\ph^\rho^\sigma^\alpha^\beta}
+ C^{\mu\nu}K\indices{_\mu_\nu^\rho^\sigma^\alpha^\beta}
= 0 \ ,
\end{equation}
the scalar equation of motion resulting from \cref{eq:transformation_scalar_row} is free from second-derivatives of the metric perturbation. Note that a side effect is the change of all the other terms of $\delta\EOM_\ph$.

From the matrix point of view, the transformations~\eqref{eq:transformation_scalar_row}, \eqref{eq:transformation_tensor_row} are operations on the \emph{rows} of $\kineticmatrix^{\alpha\beta}, \amplitudematrix^{\alpha}$, because they consist in linear combinations of $\delta\EOM_\ph, \delta\EOM_{\mu\nu}$ without mixing the variables~$\dph, \hh_{\mu\nu}$.

\subsubsection{Eliminating second derivatives of \texorpdfstring{$\dph$}{delta phi} from \texorpdfstring{$\delta\EOM_{\mu\nu}$}{delta Emunu}}
\label{subsubsec:kinetically_tensor_EOM}

The second step of the diagonalization process aims to get rid of the scalar kinetic terms in the tensor equation of motion, i.e.~to remove the off-diagonal block~$K\indices{_\mu_\nu^\ph^\alpha^\beta}$.

Unlike the previous step, this operation cannot be achieved with a mere combination of the equations of motion $\delta\EOM_\ph, \delta\EOM_{\mu\nu}$, but instead requires to combine their variables~$\dph, \hh_{\mu\nu}$ to get eigenfunctions of the system. Specifically, we introduce the \emph{eigentensor perturbation}
\begin{empheq}[box=\fbox]{equation}
\tp_{\mu\nu} \define \hh_{\mu\nu} + \hat{C}_{\mu\nu} \dph \ , \label{eq:gamma_def}
\end{empheq}
where
\begin{equation}
\hat{C}_{\mu\nu}
= \bG_4^{-1}\pa{\bG_{3,X}\bph_{,\mu}\bph_{,\nu}-\bG_{4,\ph}\bg_{\mu\nu}} \,,
\end{equation}
is the trace-reversed counterpart\footnote{Had we defined the matrices $\kineticmatrix^{\alpha\beta}, \amplitudematrix^\alpha, \massmatrix$ in terms of $h_{\mu\nu}$ instead of $\hh_{\mu\nu}$, the first step would have featured $\hat{C}_{\mu\nu}$ instead of $C_{\mu\nu}$.} of the tensor~$C_{\mu\nu}$ defined in \cref{eq:definition_C}.\footnote{Note that \cref{eq:gamma_def} agrees with the results of Ref.~\cite{Ezquiaga:2020dao}, which coincidentally was submitted on the same day as the present paper.} The fact that the same tensor appears in both diagonalization operations surely is not a coincidence, but we could not identify its fundamental origin. 

As shown in \cref{subsubsec:diagonalization_operations}, the transformation
\begin{align} 
\dph &\mapsto \dph \ , \\
\label{eq:transformation_tensor_column}
\hh_{\mu\nu} &\mapsto \tp_{\mu\nu} \ ,
\end{align}
which may be seen as an operation on the \emph{columns} of the matrix system, terminates the diagonalization procedure by removing second derivatives of the scalar perturbation from the tensor equation of motion. Its success is due to the relation
\begin{equation}
K\indices{_\mu_\nu^\ph^\alpha^\beta}
-
K\indices{_\mu_\nu^\rho^\sigma^\alpha^\beta} \hat{C}_{\rho\sigma}
= 0 \ ,
\end{equation}
between the original blocks of the kinetic matrix~$\kineticmatrix^{\alpha\beta}$. Just like in the first step, the transformation~\eqref{eq:transformation_tensor_column} modifies almost all the other blocks of $\kineticmatrix^{\alpha\beta}, \amplitudematrix^\alpha$.

In the end, our two diagonalization steps are equivalent to the following operations on the kinetic and amplitude matrices:
\begin{widetext}
\begin{align}
\kineticmatrix^{\alpha\beta}
&\mapsto \diag{\kineticmatrix}^{\alpha\beta}
= 
\begin{pmatrix}
1 & C^{\mu\nu}\\
0 & 1
\end{pmatrix}
\kineticmatrix^{\alpha\beta}
\begin{pmatrix}
1 & 0\\
-\hat{C}_{\rho\sigma} & 1
\end{pmatrix}
=
\begin{pmatrix}
\diag{K}\indices{_\ph^\ph^\alpha^\beta} &
0 \\
0 &
\diag{K}\indices{_\mu_\nu^\rho^\sigma^\alpha^\beta}
\end{pmatrix} ,
\\
\amplitudematrix^{\alpha}
&\mapsto
\diag{\amplitudematrix}^{\alpha}
=
\begin{pmatrix}
1 & C^{\mu\nu}\\
0 & 1
\end{pmatrix}
\pac{
    \amplitudematrix^{\alpha}
    +
    2\kineticmatrix^{(\alpha\beta)}
    \bar{\nabla}_\beta
    }
\begin{pmatrix}
1 & 0\\
-\hat{C}_{\rho\sigma} & 1
\end{pmatrix}
=
\begin{pmatrix}
\diag{A}\indices{_\ph^\ph^\alpha} &
\diag{A}\indices{_\ph^\rho^\sigma^\alpha} \\
\diag{A}\indices{_\mu_\nu^\ph^\alpha} &
\diag{A}\indices{_\mu_\nu^\rho^\sigma^\alpha}
\end{pmatrix} ,
\end{align}
\end{widetext}
where a slash indicates a quantity obtained after the diagonalization process. Because they are kinetically decoupled, the variables $\dph, \tp_{\mu\nu}$ must be considered the \emph{true degrees of freedom} of the linear theory.

\subsubsection{Harmonic gauge}
\label{subsubsec:harmonic}

The expressions of the matrices $\diag{\kineticmatrix}^{\alpha\beta}$ and $\diag{\amplitudematrix}^\alpha$ can be further simplified by taking advantage of the theory's \emph{gauge freedom}, stemming from the diffeomorphism invariance of the action~\eqref{eq:Action}. Under any infinitesimal transformation $x^\mu \mapsto \tilde{x}^\mu=x^\mu - \xi^\mu$, where $\xi_\mu$ is an infinitesimal vector field, the functional expression of the background fields~$\bg_{\mu\nu}, \bph$ are left unchanged if the following transformations are applied to the perturbations:
\begin{align}
h_{\mu\nu} &\mapsto \tilde{h}_{\mu\nu} = h_{\mu\nu} + 2 \xi_{(\mu;\nu)}\, \\
\dph &\mapsto \delta\tilde{\ph} = \dph + \bph_{,\rho}\xi^\rho \ .
\end{align}
It follows in particular that
\begin{equation}
\tensor{\tp}{_\mu _\nu ^;^\nu}
\mapsto
\tensor{\tilde{\tp}}{_\mu _\nu ^;^\nu}
=
\tensor{\tp}{_\mu _\nu ^;^\nu} + \Box \xi_\mu + \hat{C}_{\mu\nu} \bph_{,\rho} \xi^{\rho;\nu} \ ,
\end{equation}
up to negligible masslike terms. Since $\xi^\mu$ is arbitrary, we are free to impose the \emph{generalized harmonic gauge}
\begin{empheq}[box=\fbox]{equation}
\label{eq:harmonic_gauge}
\tensor{\tp}{^\mu^\nu_;_\nu} = 0 \ ,
\end{empheq}
because if $\tp_{\mu\nu}$ did not satisfy the above, then we could always find a gauge field that is a solution of the hyperbolic partial differential equation $\Box \xi_\mu + \hat{C}_{\mu\nu} \bph_{,\rho} \xi^{\rho;\nu} = -\tensor{\tp}{_\mu _\nu ^;^\nu}$, so that the gauge-transformed $\tilde{\tp}_{\mu\nu}$ would.

The main advantage of \cref{eq:harmonic_gauge} is that it elegantly reduces the kinetic term of $\tp_{\mu\nu}$ to
\begin{equation}
\diag{K}\indices{_\mu_\nu^\rho^\sigma^\alpha^\beta}\tp_{\rho\sigma;\alpha\beta}
= -\frac{1}{2} \bG_4 \Box \tp_{\mu\nu} \ .
\end{equation}
The other blocks of the kinetic and amplitude matrices, after diagonalization and gauge fixing, can be found in \cref{subsubsec:kinetic_matrix_diagonal,subsubsec:amplitude_matrix_diagonal}, respectively.

\section{Scalar and tensor waves}
\label{sec:scalar_tensor_waves}

Having identified the kinetically decoupled degrees of freedom  $\dph, \tp_{\mu\nu}$, we shall now focus more specifically on the case where such perturbations are propagating waves. The wave ans\"atze are presented in \cref{subsec:wave_ansatze}, where we also justify why masslike terms were dropped in the previous section. The dispersion relations of scalar and tensor waves are discussed in \cref{subsec:dispersion_relations}, and their effect on matter in \cref{subsec:polarization}.

\subsection{Wave ans\"atze and eikonal approximation}
\label{subsec:wave_ansatze}

We consider scalar and tensor perturbations under the form
\begin{align}
\label{eq:ansatz_scalar}
\dph &= \Phi \, \ex{\i v} \ , \\
\label{eq:ansatz_tensor}
\tp_{\mu\nu} &= \ta_{\mu\nu} \, \ex{\i w} \ ,
\end{align}
where $\Phi, \ta_{\mu\nu} \in \mathbb{C}$ represent the complex amplitudes of the waves, $v, w \in \mathbb{R}$ denote their respective phases.

A key assumption in this article is that the waves satisfy the \emph{eikonal} (or WKB) approximation.\footnote{Although the eikonal approximation is frequently considered a synonym of geometric optics (including in our own previous work), they are in fact not equivalent. Geometric optics consists in neglecting wave-optics effects, such as interference and diffraction. Yet such phenomena are usually studied in the framework of the eikonal approximation (see e.g. Ref.~\cite{Ezquiaga:2020spg} for a recent example in gravitational-wave physics). Recent attempts to go beyond the eikonal approximation can be found in Refs.~\cite{Harte:2018wni, Cusin:2019rmt}.} This means that the typical evolution scale of the waves' phases~$v, w$, be it temporal or spatial, is much shorter than any other characteristic length or time scale of the system. In particular, the phases are varying much quicker than the amplitudes
\begin{equation}
\partial v, \partial w \gg \partial\ln|\Phi|, \partial\ln|\ta_{\mu\nu}| \ ,
\end{equation}
which is the traditional content of the eikonal regime; they also vary much more quickly than the background fields
\begin{equation}
\partial v, \partial w \gg \partial\ln|\bph|, \partial\ln|\bg_{\mu\nu}| \ .
\end{equation}

In that context, a useful book-keeping parameter is the angular frequency of the waves,\footnote{The frequency of scalar and tensor waves could in principle be different. However, in practice we expect both types of waves to be emitted by the same kind of events, e.g., merging binaries of compact objects, and thereby with the same frequency. When only one of these waves is emitted and decays into the other sector, both are also expected to have the same frequency.} $\omega\sim\partial v, \partial w$. The many terms involved in the equations of motion can thereby be sorted depending on their power of $\omega$, i.e., depending on how many derivatives are hitting the phases~$v, w$. This implies the following hierarchy in Eq.~\eqref{eq:linearized_EoMs_formal}
\begin{equation}
\underbrace{
            \kineticmatrix^{\alpha\beta}
            \bar{\nabla}_\alpha\bar{\nabla}_\beta
            }_{\order(\omega^2)}
\gg
\underbrace{
            \amplitudematrix^{\alpha}
            \bar{\nabla}_\alpha
            }_{\order(\omega^1)}
\gg
\underbrace{
            \massmatrix
            }_{\order(\omega^0)} \ .
\end{equation}
In practice, we only keep the $\order(\omega^2)$ and $\order(\omega^1)$ terms, which rule the dispersion relations and the amplitudes of the waves, respectively. This explains why we have chosen to drop the masslike terms right from the beginning.

Note that derivatives of $\dph, \tp_{\mu\nu}$ actually contain terms with different powers of $\omega$. From the ans\"atze~\eqref{eq:ansatz_scalar}, \eqref{eq:ansatz_tensor}, we find
\begin{align}
\dph_{,\alpha}
&= \pa{\i\Phi\,q_\alpha + \Phi_{,\alpha}} \ex{\i v} \ ,
\\
\tp_{\mu\nu;\alpha}
&= \pa{\i \ta_{\mu\nu} k_\alpha + \ta_{\mu\nu;\alpha}} \ex{\i w} \ ,
\end{align}
where $q_\alpha\define v_{,\alpha}$ and $k_\alpha\define w_{,\alpha}$ are the wavevectors of the scalar and tensor waves, respectively. Any occurrence of $k_\alpha, q_\alpha$ counting as a power of $\omega$, the above expressions contain both $\order(\omega^0)$ and $\order(\omega^1)$ terms. As for the second derivatives,
\begin{align}
\dph_{;\alpha\beta}
&= \big[- \Phi\, q_\alpha q_\beta
        + 2\i\Phi_{,(\alpha} q_{\beta)}
        + \i\Phi\, q_{\alpha;\beta}
    \nonumber\\ &\qquad
        + \Phi_{;\alpha\beta}
    \big] \ex{\i v}  \ ,
\\
\tp_{\rho\sigma;\alpha\beta}
&= \big[- \ta_{\rho\sigma} k_\alpha k_\beta
        + 2\i \ta_{\rho\sigma;(\alpha}k_{\beta)}
        + \i \ta_{\rho\sigma} k_{\alpha;\beta}
     \nonumber\\ &\qquad
        + \ta_{\rho\sigma;\alpha\beta}
    \big] \ex{\i w} \ ,
\end{align}
both contain $\order(\omega^2), \order(\omega^1)$, and $\order(\omega^0)$ terms. 

\subsection{Dispersion relations}
\label{subsec:dispersion_relations}

Isolating the $\order(\omega^2)$ terms in the equations of motion, which can only come from the kinetic matrix, we find the \emph{dispersion relations} of the scalar and tensor waves,
\begin{empheq}[box=\fbox]{align}
\label{eq:dispersion_scalar_general}
\diag{K}\indices{_\ph^\ph^\alpha^\beta} q_\alpha q_\beta &= 0 \ ,
\\
\label{eq:dispersion_tensor_general}
\diag{K}\indices{_\mu_\nu^\rho^\sigma^\alpha^\beta} \ta_{\rho\sigma} k_\alpha k_\beta &= 0 \ .
\end{empheq}

\subsubsection{Tensor waves are luminal}

Let us start with the easiest of the two dispersion relations, namely the tensor one. Due to the extremely simple form of $\diag{K}\indices{_\mu_\nu^\rho^\sigma^\alpha^\beta}\propto \delta^{(\rho}_\mu\delta^{\sigma)}_\nu\bg^{\alpha\beta}$, \cref{eq:dispersion_tensor_general} actually reduces to
\begin{equation}
\label{eq:dispersion_tensor_simple}
k^\alpha k_\alpha = 0 \ ,
\end{equation}
which means that tensor waves propagate at the speed of light. It also implies that tensor waves follow null geodesics, because
\begin{equation}
0 = (k^\alpha k_\alpha)_{;\beta}
=2k^\alpha w_{;\alpha\beta}
=2k^\alpha w_{;\beta\alpha}
=2k^\alpha k_{\beta;\alpha} \ ,
\end{equation}
which is the geodesic equation. These statements are, in particular, independent of the polarization of the wave. This would not happen if the indices of $K\indices{_\mu_\nu^\rho^\sigma^\alpha^\beta}$ were intertwined in a more complicated way, as it is the case for more general Horndeski theories, such as quartic or quintic Galileons~\cite{Ezquiaga:2020dao}.

The polarization-independence of the tensor wave's dispersion relation justifies, a posteriori, the fact that we considered a single phase factor~$\ex{\i w}$ in the ansatz~\eqref{eq:transformation_tensor_row} for $\tp_{\mu\nu}$. Indeed, if the dispersion relation depended on the polarization, then each component of $\tp_{\mu\nu}$ would generally propagate at its own speed, and hence should be equipped with its own phase~$w_{\mu\nu}$.

\subsubsection{The scalar wave's velocity is inhomogeneous and anisotropic}
\label{subsubsec:dispersion_scalar}

The dispersion relation of scalar waves~\eqref{eq:dispersion_scalar_general} is phenomenologically richer. Although there is no polarization dependence by definition, the speed of scalar waves generally depends on their location and on their direction of propagation. Let us be more specific; since the tensor $\diag{K}\indices{_\ph^\ph^\alpha^\beta}$ is symmetric and real-valued, there exists a tetrad $(e^\alpha_0, \ldots, e_3^\alpha)$ such that
\begin{equation}\label{eq:3_velocities}
\diag{K}\indices{_\ph^\ph^\alpha^\beta}
\propto -e_0^\alpha e_0^\beta + c_1^2 e_1^\alpha e_1^\beta + c_2^2 e_2^\alpha e_2^\beta + c_3^2 e_3^\alpha e_3^\beta \ ,
\end{equation}
where $c_1, c_2, c_3$ are the three main phase velocities of scalar waves, as measured in the frame defined by $e_0^\mu$. In \cref{eq:3_velocities}, we have assumed that $\diag{K}\indices{_\ph^\ph^\alpha^\beta}$ does not depart too much from $\bg^{\alpha\beta}$, in the sense that its nonstandard terms are not large enough to change the causal structure of the scalar dynamics; if it were the case the theory would suffer from severe instabilities. Let $(q^\alpha)=(\omega, q\vect{u})$ be the tetrad components of the scalar wave's four-vector, with $\vect{u}=(u_1, u_2, u_3)$ a Euclidean unit vector. Then the local scalar phase velocity reads
\begin{equation}
c\e{S}
\define \frac{\omega}{q}
= \sqrt{(c_1 u_1)^2 + (c_2 u_2)^2  + (c_3 u_3)^2} \ ,
\end{equation}
in the rest frame defined by $e_0^\mu$.

From the expression of $\diag{K}\indices{_\ph^\ph^\alpha^\beta}$ provided in \cref{subsubsec:kinetic_matrix_diagonal}, we can see that its failure to be proportional to $\bg^{\alpha\beta}$, i.e., the departure from $c\e{S}=1$, is due to the coupling functions $\bG_{2,XX}, \bG_{3,X}, \bG_{3,X\ph}, \bG_{3,XX}$. In the limit where such departures are small, we have\footnote{While \cref{eq:small_deviations} may also hold for superluminal scalar waves, we refrain from considering that case because its interpretation would be unclear.} 
\begin{multline} \label{eq:small_deviations}
\bG_{2,X} |c\e{S}-1|
= \order\Big[
            \bG_{2,XX}(\partial\bph)^2,
            \bG_{3,X}\partial^2\bph,
            \bG_{3,\ph X}(\partial\bph)^2, \\
            \bG_{3,XX}(\partial\bph)^2\partial^2\bph
        \Big] \ .
\end{multline}
Conversely, if we exclude any background fine tuning, then
\begin{equation}\label{eq:luminal_scalar_wave_condition}
c\e{S}=1
\Longleftrightarrow
G_{2,XX}=G_{3,X}=0 \ .
\end{equation}
In other words, scalar waves are luminal if and only if $\ph$ is a conformally-coupled quintessence field, in agreement with the findings of Ref.~\cite{Garoffolo:2019mna}.

\subsection{The effect of scalar and tensor waves on matter}
\label{subsec:polarization}

Consider the superposition of a scalar and a tensor wave. What is their effect on the matter through which they propagate, and how can they be detected?

\subsubsection{Observables are curvature perturbations}

In the action~\eqref{eq:Action}, matter is coupled to the spacetime geometry only; in particular it is not directly coupled to the scalar field. Therefore, observable effects of the scalar and tensor perturbations must be looked for in the \emph{spacetime curvature}, i.e., in the Riemann tensor. The perturbation of the latter around its background value $\bar{R}_{\mu\nu\rho\sigma}$ reads, at linear order~\cite{Straumann:2013spu},
\begin{equation}
\label{eq:curvature_perturbation}
\delta R_{\mu\nu\rho\sigma}
=
\frac{1}{2} \pa{
                h_{\mu\sigma;\nu\rho}
                - h_{\mu\rho;\nu\sigma}
                - h_{\nu\sigma;\mu\rho}
                + h_{\nu\rho;\mu\sigma}
                } ,
\end{equation}
and thereby depends on the original metric perturbation
\begin{align}
\label{eq:decomposition_metric_perturbation}
h_{\mu\nu}
= \hat{\tp}_{\mu\nu} - C_{\mu\nu}\dph
\define h_{\mu\nu}\h{T} + h_{\mu\nu}\h{S} \ .
\end{align}
In other words, the curvature perturbation generally picks up two distinct contributions: a rather standard one from the tensor wave via $h_{\mu\nu}\h{T}=\hat{\tp}_{\mu\nu}$, but also one from the scalar wave via $h_{\mu\nu}\h{S}=-C_{\mu\nu}\delta\ph$. \emph{This is how the fifth force associated with scalar waves arises in the Jordan frame}. The above emphasizes the importance of a clear identification of the theory's degrees of freedom, and notably $\tp_{\mu\nu}\not= h_{\mu\nu}$.

To be more explicit, by combining \cref{eq:curvature_perturbation,eq:decomposition_metric_perturbation} we see that the curvature perturbation can be written as the superposition of two waves, respectively linked to $h_{\mu\nu}\h{T}$ and $h_{\mu\nu}\h{S}$,
\begin{empheq}[box=\fbox]{equation}
\delta R_{\mu\nu\rho\sigma}
= \mathcal{R}\h{T}_{\mu\nu\rho\sigma} \ex{\i w}
    + \mathcal{R}\h{S}_{\mu\nu\rho\sigma} \ex{\i v}
\end{empheq}
and whose amplitudes read, at leading order in $\omega$,
\begin{align}
\mathcal{R}\h{T}_{\mu\nu\rho\sigma}
&= 2k_{[\nu}H\h{T}_{\mu][\rho} k_{\sigma]}
= 2k_{[\nu}\hat{\ta}_{\mu][\rho} k_{\sigma]} \ ,
\\
\label{eq:amplitude_scalar_curvature}
\mathcal{R}\h{S}_{\mu\nu\rho\sigma}
&= 2q_{[\nu}H\h{S}_{\mu][\rho} q_{\sigma]}
= -2\Phi q_{[\nu}C_{\mu][\rho} q_{\sigma]} \ .
\end{align}

\subsubsection{The tensor contribution is standard}
\label{subsubsec:tensor_is_standard}

The contribution of the tensor wave to the metric (and curvature) perturbation is what one usually refers to as a GW. Its properties were analyzed in details in our earlier work~\cite{Dalang:2019rke}. Let us briefly summarize its main findings in the following.

Because of the harmonic-gauge condition~\eqref{eq:harmonic_gauge}, which at leading order imposes $k^\nu \ta_{\mu\nu}=0$ on the amplitude of $\tp_{\mu\nu}$, it can be shown that its trace-reversed counterpart is decomposed into a gauge mode and a transverse-traceless mode as
\begin{equation}
\hat{\ta}_{\mu\nu} \define H_{\mu\nu}\h{T} = H_{\mu\nu}\h{G} + H_{\mu\nu}\h{TT} \ .
\end{equation}
The gauge mode, which takes the form $H_{\mu\nu}\h{G}=2 k_{(\mu}H_{\nu)}$ where $H_\nu$ is a vector field, is nonphysical: it does not contribute to the curvature perturbation, it does not carry energy-momentum, and it can always be locally removed by a gauge transformation.

The transverse-traceless mode~$H_{\mu\nu}\h{TT}$ contains the physics of the tensor wave. Its expression is conveniently written in terms of a null tetrad $(k_\mu, n_\mu, m_\mu, m^*_\mu)$, whose vectors are all null, and where a star denotes a complex conjugate; the only nonzero inner products of the tetrad are $\bg^{\mu\nu} m_\mu m_\nu^* = \bg^{\mu\nu} k_\mu n_\nu =1$. The vectors $m_\mu, m^*_\mu$ can be seen as spanning a spatial screen that is orthogonal to the wave's direction of propagation (see Ref.~\cite{Dalang:2019rke} for details). The transverse-traceless mode then reads
\begin{align}
H_{\mu\nu}\h{TT}
= H_\circlearrowleft m_\mu m_\nu
    + H_\circlearrowright m^*_\mu m^*_\nu \ ,
\end{align}
thereby defining the complex amplitudes~$H_\circlearrowleft, H_\circlearrowright$ of the left-handed and right-handed helicity modes. These are related to the usual plus and cross polarizations~$H_+, H_\times$ through
\begin{align}
H_\circlearrowleft
&= H_+ - \i H_\times \ ,
\\
H_\circlearrowright
&= H_+ + \i H_\times .
\end{align}

In the rest frame of any observer, if the tensor wave propagates in the $z$-direction, then the amplitude of the associated curvature perturbation reads
\begin{equation}
\label{eq:tidal_forces_scalar_tensor}
(\mathcal{R}_{0i0j}\h{T})
=
\frac{\omega\e{T}^2}{2}
\begin{pmatrix}
H_+ & H_\times & 0 \\
H_\times & -H_+ & 0 \\
0 & 0 & 0 
\end{pmatrix}
\ ,
\end{equation}
where $\omega\e{T}$ is the observed cyclic frequency of the tensor wave. No force is produced in the direction of propagation.

\subsubsection{Effect of a luminal scalar wave}
\label{subsubsec:effect_luminal_scalar_wave}

The tidal forces provoked by scalar waves depend on their propagation speed. Let us start with the luminal case ($c\e{S}=1$). Following the discussion of \cref{subsubsec:dispersion_scalar}, if we do not allow for fine-tuned setups, then the luminal condition imposes $G_{3,X}=0$, so that $C_{\mu\nu}=\bG_{4}^{-1}\bG_{4,\ph}\bg_{\mu\nu}$. It is then straightforward to show that, in the rest frame of any observer, the associated curvature perturbation reads
\begin{equation}
\label{eq:tidal_forces_scalar_luminal}
(\mathcal{R}_{0i0j}\h{S})
=
-\frac{\omega\e{S}^2}{2} \frac{\bG_{4,\ph}}{\bG_4}
\begin{pmatrix}
\Phi & 0 & 0 \\
0 & \Phi & 0 \\
0 & 0 & 0 
\end{pmatrix} \ ,
\end{equation}
where, again, the $z$-direction corresponds to the local direction of propagation of the scalar wave, and $\omega\e{S}$ is the observed cyclic frequency of the scalar wave.

\Cref{eq:tidal_forces_scalar_luminal} represents the tidal forces that are intuitively expected from a scalar wave. If a ring of freely-falling particles were placed in the $xy$-plane, then the ring's radius would periodically increase and decrease by an amount proportional to $\Phi$. It is, however, interesting to notice that the wave's effects remain transverse, in the sense that there are no tidal forces along the direction of propagation. If $G_{4,\ph}=0$, the scalar wave decouples and travels without interacting with interferometers.

\subsubsection{Effect of a subluminal scalar wave}

The phenomenology of subluminal scalar waves ($c\e{S}<1$) is richer. In any observer's rest frame, the amplitude~\eqref{eq:amplitude_scalar_curvature} of the curvature perturbation that it causes reads
\begin{equation}
\mathcal{R}_{0i0j}\h{S}
= -\frac{1}{2} \, \Phi
    \pac{
        q_i q_j C_{00}
        - 2 q_{(i} C_{j)0}
        + \omega\e{S}^2 C_{ij}
        } ,
\end{equation}
which now features two unrelated directions: the wavevector~$\vect{q}$ on the one hand, and the background scalar field's gradient~$\vect{\nabla}\bph$ present in $C_{ij}$ on the other hand. As a consequence, tidal forces are generally triaxial; in particular, they are no longer transverse. More explicitly, since $\mathcal{R}_{0i0j}\h{S}$ is symmetric and real-valued, it may be diagonalized in an orthonormal system $(\vect{e}_a)_{a=1,\ldots,3}$ as
\begin{equation}
\label{eq:tidal_forces_scalar_subluminal}
(\mathcal{R}_{0a0b}\h{S})
=
-\frac{\omega\e{S}^2}{2}
\begin{pmatrix}
\mathcal{C}_1\Phi & 0 & 0 \\
0 & \mathcal{C}_2\Phi & 0 \\
0 & 0 & \mathcal{C}_3\Phi
\end{pmatrix} \ ,
\end{equation}
where $\mathcal{C}_1, \mathcal{C}_2, \mathcal{C}_3$ are three dimensionless shape parameters, which depend on $\bG_{3,X}/\bG_{4}$, $c\e{S}$, the derivatives of $\bph$, and the angle between $\vect{q}$ and $\vect{\nabla}\bph$.

Note that the orthonormal basis $(\vect{e}_a)$ used in \cref{eq:tidal_forces_scalar_subluminal} is generally different from the orthonormal basis $(\vect{e}_i)$ used in \cref{eq:tidal_forces_scalar_tensor,eq:tidal_forces_scalar_luminal}. In particular, the direction $a=3$ does not always coincide with the direction of propagation of the wave.

In a scenario of chameleon screening, we may expect the derivatives $\partial\bph$ to be suppressed at the observer's location. This would imply $c\e{S}\approx 1$ and $C_{\mu\nu}\propto \bg_{\mu\nu}$, thereby bringing us back to \cref{subsubsec:effect_luminal_scalar_wave}. Hence, screening would not remove the effect of a scalar wave, but rather reduce it to that of a luminal wave which would be delayed with respect to the tensor wave.

\section{(Non-)interaction between scalar and tensor waves}
\label{sec:Non_Interactions}

In the previous sections, we have shown that in reduced Horndeski theories, GWs consist of the superposition of a tensor wave (two degrees of freedom) and a scalar wave (one degree of freedom). While tensor waves propagate at the speed of light, scalar waves do not in general. In this section, we investigate the evolution of the amplitude of tensor and scalar waves, and their mutual interactions.

\subsection{Evolution of the amplitudes}
\label{subsec:evolution_amplitudes}

The standard procedure to get evolution equations for the wave amplitudes consists in extracting the $\order(\omega^1)$ terms in the equations of motion. This is easily understood starting with the tensor wave, whose kinetic term reads
\begin{align}
\diag{K}\indices{_\mu_\nu^\rho^\sigma^\alpha^\beta}
\tp_{\rho\sigma;\alpha\beta}
&= - \frac{1}{2} \bG_4 \Box \tp_{\mu\nu}
\\
&= \frac{1}{2} \bG_4 \pa{\ta_{\mu\nu}k_\alpha k^\alpha - \i\propagator \ta_{\mu\nu}} \ex{\i w} \ ,
\end{align}
up to $\order(\omega^0)$ terms, and with the differential operator
\begin{equation}
\propagator \define 2 k^\alpha \bar{\nabla}_\alpha + \tensor{k}{^\alpha_;_\alpha} \,.
\end{equation}
The $\order(\omega^1)$ term in the above contains $\propagator \ta_{\mu\nu}$, whose function is to propagate $\ta_{\mu\nu}$ in the direction of $k^\alpha$, i.e.~along the null geodesic followed by the tensor wave. The other $\order(\omega^1)$ contributions to $\delta\EOM_{\mu\nu}$ come from the amplitude matrix, which encodes both self interactions (diagonal terms), and interactions with the scalar waves (off-diagonal terms). In GR, the result would simply read $\propagator\ta_{\mu\nu}=0$, leading to the fact that the GW amplitude essentially decreases as the area of its wavefront. Here, we have instead
\begin{equation}
\label{eq:amplitude_evolution_tensor}
0 = - \frac{\bG_4}{2} \propagator \ta_{\mu\nu}
    +  \diag{A}\indices{_\mu _\nu ^\rho ^\sigma ^\alpha}
        \ta_{\rho \sigma} k_\alpha
    + \pa{
        \diag{A}\indices{_\mu _\nu ^\varphi ^\alpha}  q_\alpha \Phi
        } \ex{\i (v-w)} \ .
\end{equation}
The same procedure, i.e.~extracting $\order(\omega^1)$ terms, applied to the scalar equation of motion yields
\begin{gather}
\label{eq:amplitude_evolution_scalar}
0 = \propagator\e{S}\Phi
        + \diag{A}\indices{_\ph ^\ph ^\alpha}
        q_\alpha \Phi
    + \pa{
        \diag{A}\indices{_\ph^\rho^\sigma^\alpha} \ta_{\rho\sigma} k_\alpha
        } \ex{\i (w-v)} \\
\text{with} \qquad
\propagator\e{S} \define 
\diag{K}\indices{_\ph^\ph^\alpha^\beta}
\pa{
    2 q_{\alpha} \bar{\nabla}_{\beta} 
    + q_{\alpha;\beta}
    } \ . 
\end{gather}

Because scalar and tensor waves may propagate at different speeds, their phases $v$ and $w$ may differ. This makes the analysis of the combined evolution of $\Phi, \Gamma_{\mu\nu}$ more subtle than the scalar-wave-free setup studied in Ref.~\cite{Dalang:2019rke}. In the following, we shall successively consider the cases where scalar waves are luminal ($c\e{S}=1$), quasiluminal ($c\e{S}\approx 1$), and finally nonluminal ($c\e{S}\neq 1$).

\subsection{Luminal scalar waves: no interactions}
\label{subsec:no_interactions_luminal_scalar}

As discussed in \cref{subsubsec:dispersion_scalar}, the condition $c\e{S}=1$ imposes $G_{2,XX}=G_{3,X}=0$, which drastically simplifies the problem. In particular, \cref{eq:amplitude_evolution_tensor,eq:amplitude_evolution_scalar} become propagation equations for the amplitudes~$\Phi, \ta_{\mu\nu}$ only. In what follows, we assume that the scalar and tensor waves propagate along the same geodesic and have the same frequency, so that $k^\mu=q^\mu$. This assumption is motivated by the fact that we are primarily interested in the interaction of scalar and tensor waves originating from the same source. It is nevertheless straightforward to generalize our results to other setups.

\subsubsection{Tensor amplitude}
\label{subsubsec:tensor_amplitude_luminal}

First consider the evolution of the tensor amplitude, governed by \cref{eq:amplitude_evolution_tensor},  which for $v=w$ becomes
\begin{equation}
\label{eq:amplitude_evolution_tensor_luminal}
0 = - \frac{1}{2} \bG_4 \propagator \ta_{\mu\nu}
    + \diag{A}\indices{_\mu _\nu ^\rho ^\sigma ^\alpha}
        k_\alpha \ta_{\rho \sigma} 
    + \diag{A}\indices{_\mu _\nu ^\varphi ^\alpha}
        k_\alpha \Phi \ .
\end{equation}
The amplitude terms are obtained from \cref{subsec:diagonal} by taking $G_{2,XX}=G_{3,X}=0$,
\begin{align}
\diag{A}\indices{_\mu _\nu ^\rho ^\sigma ^\alpha}
        k_\alpha \ta_{\rho \sigma}
&= \bG_{4,\ph} \Big[
            2 \bph^{,\rho} \hat{\ta}_{\rho(\mu} k_{\nu)}
            - k^\rho\bph_{,\rho} \hat{\ta}_{\mu\nu} 
            \Big] ,
\\
\diag{A}\indices{_\mu _\nu ^\varphi ^\alpha} k_\alpha
&= - \pa{ \bG_{2,X} + 2\bG_{3,\ph} }
    \pac{
        \bph_{(\mu}k_{,\nu)}
        - \frac{1}{2} k^\rho\bph_{,\rho} \bg_{\mu\nu}
        }
    \nonumber\\&\quad
+ \bG_4^{-1}\bG_{4,\ph}^2
        \pac{
            \bph_{,(\mu} k_{,\nu)}
            + \frac{1}{2} k^\rho\bph_{,\rho} \bg_{\mu\nu}
            } .
\end{align}

As mentioned in \cref{subsubsec:tensor_is_standard}, the tensor amplitude can be decomposed into a nonphysical gauge mode and physical transverse-traceless mode, which in turn consists of two helicity modes with complex amplitudes
\begin{align}
H_\circlearrowleft
&= m^*_\mu m^*_\nu H\e{TT}^{\mu\nu}
= m^*_\mu m^*_\nu \hat{\ta}^{\mu\nu}
= m^*_\mu m^*_\nu \ta^{\mu\nu} \ ,
\\
H_\circlearrowright
&=m_\mu m_\nu \ta^{\mu\nu} \ .
\end{align}
Hence, the evolution of $H_\circlearrowleft$ and $H_\circlearrowright$ (or equivalently of $H_+, H_\times$) can be obtained by projecting \cref{eq:amplitude_evolution_tensor_luminal} on $m^*_\mu m^*_\nu$ and $m_\mu m_\nu$. Since $m_\mu, m^*_\mu$ are null and orthogonal to $k_\mu$, all the projected amplitude terms read
\begin{gather}
m_\mu m_\nu
\diag{A}\indices{^\mu ^\nu ^\rho ^\sigma ^\alpha}
k_\alpha \ta_{\rho \sigma}
= k^\rho\bG_{4,\rho} H_\circlearrowright
\\
m_\mu^* m_\nu^*
\diag{A}\indices{^\mu ^\nu ^\rho ^\sigma ^\alpha}
k_\alpha \ta_{\rho \sigma}
= k^\rho\bG_{4,\rho} H_\circlearrowright
\\
m_\mu m_\nu \diag{A}\indices{^\mu ^\nu ^\varphi ^\alpha} k_\alpha
=
m_\mu^* m_\nu^* \diag{A}\indices{^\mu ^\nu ^\varphi ^\alpha} k_\alpha
= 0 \ .
\end{gather}

Assuming without loss of generality that $m_\mu$ and $m^*_\mu$ are parallel-transported along the worldline of the tensor wave, the projections of \cref{eq:amplitude_evolution_tensor_luminal} on $m_\mu m_\mu$ and $m_\mu^*m_\mu^*$ then reduce to
\begin{equation}
\label{eq:propagation_tensor_luminal_result}
\propagator\pac{ \sqrt{G_4(\bph)} \, H_\ocircle } = 0 \ ,
\end{equation}
where $H_\ocircle$ stands for either of $H_\circlearrowleft, H_\circlearrowright$; but \cref{eq:propagation_tensor_luminal_result} would also apply to $H_+, H_\times$ which are linear superpositions of the circular modes. We conclude, in particular, that \emph{the physical tensor modes are mutually independent and decoupled from the scalar wave}.

\Cref{eq:propagation_tensor_luminal_result} is identical to the main result of Ref.~\cite{Dalang:2019rke} in which scalar waves were initially neglected. All the conclusions of that reference thus hold in the presence of luminal scalar waves. In particular: (i) the polarization of a tensor wave is parallel-transported along the wave's worldline; (ii) as the wave propagates, its amplitude changes as\footnote{This step uses the fact that, in the operator~$\propagator$, the wavefront's expansion rate reads $k\indices{^\mu_;_\mu}=2\dd\ln D/\dd\lambda$, where $\lambda$ is an affine parameter for the GW's geodesic. See Sec.~III.C.3 of Ref.~\cite{Dalang:2019rke} for details.}
\begin{empheq}[box=\fbox]{equation}
\label{eq:amplitude_evolution_tensor_luminal_result}
H_\ocircle \propto \frac{1}{\sqrt{G_4(\bph)}\,D(z)} \ ,
\end{empheq}
where $z$ is the \emph{observed} redshift and $D(z)=(1+z)D\e{A}$ is the corrected luminosity distance, with $D\e{A}$ the \emph{observed} angular diameter distance to the source. \Cref{eq:amplitude_evolution_tensor_luminal_result} thus applies regardless of the observer's motion, lensing, integrated Sachs-Wolfe effect, etc. Finally, (iii) since the \emph{gravitational distance}~$D\e{G}$ is extracted from GW observations in such a way that $H_\ocircle \propto (1+z)/D\e{G}$, that distance is related to the electromagnetic luminosity distance~$D\e{L}=(1+z)^2 D\e{A}$ as\footnote{Note that $D\e{G}$ is, in fact, fundamentally related to the geometric angular diameter distance $D\e{A}$. \Cref{eq:D_G-D_L} implicitly assumes the validity of the distance-duality law for electromagnetic signals, which requires the conservation of photon number.}
\begin{equation}
\label{eq:D_G-D_L}
D\e{G} =
\sqrt{ \frac{G_4(\bph\e{o})}{G_4(\bph\e{s})} } \, D\e{L} \ ,
\end{equation}
where $\bph\e{o}, \bph\e{s}$ are the values of the background scalar field at the observation and emission events. We stress that the word ``background'' does not necessarily refer to a homogeneous-isotropic cosmological setup, but rather designates the scalar field's state without scalar waves.

\subsubsection{Scalar amplitude}
\label{subsubsec:scalar_amplitude_luminal}

We then turn to the evolution of the scalar amplitude, which is dictated by \cref{eq:amplitude_evolution_scalar}. For luminal scalar waves, $\diag{K}\indices{_\ph^\ph^\alpha^\beta}\propto \bg^{\alpha\beta}$, and hence it is fully determined by its trace. Let us introduce
\begin{equation}
\label{eq:definition_N}
N \define \frac{1}{4} \, \bg_{\alpha\beta} \diag{K}\indices{_\ph^\ph^\alpha^\beta}
= \bG_{2,X} + 2\bG_{3,\ph} + 3\bG_4^{-1} \bG_{4,\ph}^2 ,
\end{equation}
in terms of which $\propagator\e{S}=N\propagator$, so that \cref{eq:amplitude_evolution_scalar} becomes
\begin{equation}
\label{eq:amplitude_evolution_scalar_luminal}
0 = N \propagator\Phi
    + \diag{A}\indices{_\ph^\ph^\alpha} k_\alpha \Phi
    + \diag{A}\indices{_\ph^\rho^\sigma^\alpha} k_\alpha \ta_{\rho\sigma}\ .
\end{equation}
The amplitude terms are obtained from \cref{subsubsec:amplitude_matrix_diagonal} by imposing $\bG_{2,XX}=\bG_{3,X}=0$, which yields
\begin{align}
\diag{A}\indices{_\ph^\ph^\alpha} k_\alpha
&= k^\rho N_{,\rho} \ ,
\\
\diag{A}\indices{_\ph^\rho^\sigma^\alpha} k_\alpha \ta_{\rho\sigma}
&= 0 \ ,
\end{align}
and hence \cref{eq:amplitude_evolution_scalar_luminal} reduces to $\propagator(N\Phi)=0$. This confirms, in particular, that \emph{there is no interaction between tensor waves and luminal scalar waves}.

Following the same logic as in the tensor-wave case, we conclude that the amplitude of a scalar wave evolves as
\begin{empheq}[box=\fbox]{equation}
\Phi \propto \frac{1}{\sqrt{N(\bph, \bX)}\, D} \ . 
\end{empheq}
We may define a notion of \emph{scalar distance} similarly to how we defined the gravitational distance for tensor waves, namely the quantity that governs the wave's amplitude as $\Phi \propto (1+z)/D\e{S}$. Following that definition, the scalar distance would read
\begin{equation}
D\e{S} = \sqrt{\frac{N(\bph\e{o}, \bX\e{o})}{N(\bph\e{s}, \bX\e{s})}} \, D\e{L} \not= D\e{G}\ ,
\end{equation}
with the function $N$ given in \cref{eq:definition_N}.

\subsection{Quasiluminal scalar waves: negligible interactions}
\label{subsec:quasiluminal_scalar_waves}

We have seen that there are no interactions between scalar and tensor waves if $c\e{S}=1$. Does that conclusion hold when $c\e{S}$ is close enough to $1$, so that \cref{eq:amplitude_evolution_tensor,eq:amplitude_evolution_scalar} can be used to study the evolution of $\Phi, \ta_{\mu\nu}$? In other words, is the problem continuous in the limit $c\e{S}\to 1$?

In order to address that question, we shall first determine the condition on $c\e{S}$ such that $v\approx w$. Consider for simplicity that a scalar wave and a tensor wave are emitted simultaneously, and that they are initially in phase.\footnote{This does not restrict the generality of the discussion, since any initial relative phase can always be absorbed in the complex amplitudes $\Phi, \ta_{\mu\nu}$.} Let us determine the phase drift~$v-w$ along the tensor wave's worldline, which we affinely parametrize with $\lambda$. The phase drift from emission ($\lambda\e{s}$) to observation ($\lambda\e{o}$) then reads
\begin{align}
v - w
&= \int_{\lambda\e{s}}^{\lambda\e{o}} \dd \lambda
    \pa{\frac{\dd v}{\dd \lambda}-\ddf{w}{\lambda}}
\\
&= \int_{\lambda\e{s}}^{\lambda\e{o}} \dd \lambda
    \pa{\partial_\mu v-\partial_\mu w} \frac{\dd x^\mu}{\dd \lambda}
\\
&= \int_{\lambda\e{s}}^{\lambda\e{o}} \dd \lambda \; (q_\mu-k_\mu) k^\mu
\\
&= \int_{\lambda\e{s}}^{\lambda\e{o}} \dd \lambda \; \omega\e{T}\omega\e{S} \, \frac{1-c\e{S}}{c\e{S}} \ ,
\end{align}
where we decomposed the wave four-vectors on a tetrad so as to exhibit their frequencies and phase velocities, $k^\mu=\omega\e{T}(1,0,0,1)$ and $q^\mu=\omega\e{S}(1,0,0,1/c\e{S})$; we neglected the relative deflection of scalar and tensor waves for simplicity. In the rest frame associated with the tetrad, the product $\dd\ell=\omega\e{T}\dd\lambda$ represents the physical distance over which the tensor wave travels as the affine parameter changes by $\dd\lambda$. Therefore,
\begin{align}\label{eq:phase_condtion}
v-w
= \int_0^{\ell\e{s}} \dd\ell\; \omega\e{S} \, \frac{1-c\e{S}}{c\e{S}} 
\define \ell\e{s} \ev{ \omega\e{S} \, \frac{1-c\e{S}}{c\e{S}} } .
\end{align}
where $\ell\e{s}$ is the affine-parameter based distance between the source and the observer, and $\ev{\ldots}$ is the $\ell$-weighted average.

We conclude that the difference between the phases $v$ and $w$ in \cref{eq:amplitude_evolution_tensor,eq:amplitude_evolution_scalar} is negligible if
\begin{equation}
\label{eq:phase_velocity_condition}
\ell\e{s} \ev{ \omega\e{S} \, \frac{1-c\e{S}}{c\e{S}} } \ll 1 \ .
\end{equation}
This condition could have been guessed from intuitive arguments. Two waves can be considered to stay in phase if the delay of the slowest wave with respect to the quickest remains much smaller than their period. If the waves are emitted simultaneously and travel over a distance $\ell\e{s}$, then the delay between the receptions of scalar and tensor waves is on the order of  $(1-c\e{S})\ell\e{s}/c\e{S}$. The condition for that delay to remain much smaller than $T\e{S}=2\pi/\omega\e{S}$ thus matches \cref{eq:phase_velocity_condition}.

Relaxing the condition $c\e{S}=1$, and hence allowing $G_{2,XX}, G_{3,X}\neq 0$, is comparable to opening Pandora's box and pouring its content into \cref{eq:amplitude_evolution_tensor,eq:amplitude_evolution_scalar}. Let us respectively denote $\Delta_\Phi, \Delta_\ocircle$ the sum of these new terms in the evolution equations for the scalar and tensor amplitudes,
\begin{align}
\propagator\pac{\sqrt{N(\bph, \bX)} \,\Phi} &= \Delta_\Phi \ ,
\\
\propagator\pac{\sqrt{G_4(\bph)}\, H_\ocircle} &= \Delta_\ocircle \ .
\end{align}
For instance, $\Delta_\Phi$ contains terms such as $\bG_{3,X}\Box\bph \propagator\Phi$ or $\bG_{2,XX}\bX^{,\alpha} q_\alpha\Phi$, and similarly for $\Delta_\ocircle$. In fact, careful examination of the terms composing $\Delta_\Phi, \Delta_\ocircle$ reveals that they are all similar to the ones present in $|c\e{S}-1|$, as listed in \cref{eq:small_deviations}. To be more specific, we have
\begin{equation}
\Delta_\Phi, \Delta_\ocircle \sim |c\e{S}-1|\times(\omega\partial\bph) \ .
\end{equation}
But \cref{eq:phase_condtion} shows that if $v-w\ll 1$, then $|c\e{S}-1|\ll\order(\omega^{-1})$, which implies that $\Delta_\Phi, \Delta_\ocircle$ are actually smaller than $\order(\omega^0)$, i.e., smaller than masslike terms within the eikonal hierarchy.

We conclude that, for quasiluminal scalar waves, scalar-tensor interactions can be safely neglected, so that the results of \cref{subsec:no_interactions_luminal_scalar} hold. In other words, the problem is indeed continuous in the limit $c\e{S}\to 1$.

\subsection{Nonluminal scalar waves: incoherent interference}
\label{subsec:nonluminal}

Let us finally tackle the case $c\e{S}\neq 1$ beyond \cref{subsec:quasiluminal_scalar_waves}. An important technical difficulty here is that the wavefronts and worldlines of the scalar and tensor waves generally differ. Since \cref{eq:amplitude_evolution_tensor,eq:amplitude_evolution_scalar} only tell us about the local evolution of $\Phi,\Gamma_{\mu\nu}$ along their respective worldlines, it is impossible to simultaneously solve for their evolution equations: they concern distinct lines in spacetime. Thus, we should in principle consider the wavefronts of both waves and analyze their overlap.

Fortunately, such an elaborate treatment is unnecessary if we assume that the two waves are propagating in almost the same direction. For concreteness, let us take the evolution of $\Gamma_{\mu\nu}(\lambda)$ along the null geodesic of the tensor wave. Since \cref{eq:amplitude_evolution_scalar} does not directly tell us about $\Phi(\lambda)$, we may simply consider it as an unknown function, which varies much slower than the phase drift $v(\lambda)-w(\lambda)$. This is justified by\footnote{Recall that a derivative with respect to the affine parameter counts as one power of $\omega$, since $\lambda$ is connected to proper distance~$\ell$ as $\dd\lambda=\omega^{-1}\dd\ell$ in the frame of an observer who would measures $\omega$.}
\begin{equation}
\ddf{(v-w)}{\lambda}
= \omega\e{T}\omega\e{S} \, \frac{1-c\e{S}}{c\e{S}}
= \order(\omega^2)
\gg
\ddf{\Phi}{\lambda} = \order(\omega^1) \ ,
\end{equation}
following the reasoning of \cref{subsec:quasiluminal_scalar_waves}, which requires that the spatial directions of $k^\mu$ and $q^\mu$ coincide indeed. Note that $\dd(v-w)/\dd \lambda \gg \dd\Phi/\dd\lambda$ only holds if $c\e{S}$ is not too close from $1$, which is why we treated the quasiluminal case separately.

Similarly to \cref{subsubsec:tensor_amplitude_luminal}, we may project \cref{eq:amplitude_evolution_tensor} on the helicity basis; we notice that the projected tensor self-interaction term is identical to the luminal case, and we find
\begin{equation}
\label{eq:amplitude_evolution_tensor_nonluminal}
\frac{2}{\sqrt{G_4(\bph)}D}\ddf{\sqrt{G_4(\bph)}D H_\ocircle}{\lambda}
= 
\underbrace{
\diag{A}\indices{_\ocircle ^\varphi ^\alpha}  q_\alpha \Phi \, \ex{\i (v-w)}
}_{\text{oscillates rapidly}} \ ,
\end{equation}
where $\ocircle$ can stand for either $\circlearrowleft$ or $\circlearrowright$, and $D=(1+z)D\e{A}$. Integrating~\cref{eq:amplitude_evolution_tensor_nonluminal} between~$\lambda\e{s}$ close to the source and a $\lambda$ then yields
\begin{multline}
H_\ocircle(\lambda)
= \sqrt{\frac{\bG_4(\lambda\e{s})}{\bG_4(\lambda)}}
    \frac{D(\lambda\e{s})}{D(\lambda)} \,
    H_\ocircle(\lambda\e{s})
    \\
    + \frac{1}{2}
    \int_{\lambda\e{s}}^{\lambda} \dd\lambda' \;
        \sqrt{\frac{\bG_4(\lambda')}{\bG_4(\lambda)}} 
        \frac{D(\lambda')}{D(\lambda)} \,
        \diag{A}\indices{_\ocircle ^\varphi ^\alpha}  q_\alpha \Phi \, \ex{\i (v-w)} \ ,
\end{multline}
where the first term is identical to the noninteracting case, while the second one is the average value of a rapidly oscillating function, which thus is very small. We shall call \emph{incoherent interference} this phase-related suppression of interactions.

A similar reasoning could be applied to the scalar-amplitude case, and would yield the same effective suppression of the scalar-tensor interaction due to the incoherence of the two waves. Note however that the scalar self interaction generally differs from \cref{subsubsec:scalar_amplitude_luminal}, because $\diag{A}\indices{_\ph^\ph^\alpha}$ contains many additional terms when $c\e{S}\neq 1$. Since the emission of scalar waves is 
expected to be extremely small, and since their conversion from tensor waves is suppressed by incoherent interference, we did not judge necessary to further push the present analysis, which already implies that scalar radiation is irrelevant to the tensor modes in reduced Horndeski theories.

One possible loophole in the above, however, is the assumption that scalar and tensor waves propagate (almost) in the same direction. For example, this set-up does not cover the Cherenkov-like scalar radiation that may be emitted by the passage of a tensor wave. In that case indeed, the direction of constructive interference for scalar waves would typically form an angle $\theta=\arccos(c\e{S})$ with the direction of propagation of the tensor wave. The treatment of such a case nevertheless requires the use of tools that are beyond the scope of this article, and hence left for future work.

\section{Summary and conclusion}
\label{sec:conclusion}

We have studied the propagation of GWs in the subset of Horndeski theories where tensor perturbations propagate at light speed (\emph{reduced Horndeski theories}). Unlike our previous work on the same topic~\cite{Dalang:2019rke}, we now have accounted for propagating scalar perturbations, i.e.~scalar waves, which made the present study technically much more involved.

We derived for the first time the complete set of equations governing the intertwined dynamics of scalar and tensor perturbations $\dph, h_{\mu\nu}$, for an arbitrary background scalar field and spacetime geometry~$\bph, \bg_{\mu\nu}$. Specifically, computations were made at linear order in the perturbations, and in the limit where they vary on much smaller scales than the background.

By diagonalizing the kinetic term of the resulting system of equations of motion, we have shown that the metric perturbation $h_{\mu\nu}$ is not a fundamental degree of freedom of the linearized theory. Indeed, due to nonminimal couplings, $h_{\mu\nu}$ actually encompasses some scalar information, which can be removed by considering $\tp_{\mu\nu}\define \hh_{\mu\nu}+\hat{C}_{\mu\nu}\dph$ instead, $C_{\mu\nu}$ being defined in \cref{eq:definition_C}. The quantities $\dph, \tp_{\mu\nu}$ then represent the true scalar and tensor degrees of freedom of the linearized theory.

Considering wavelike ans\"atze for $\dph, \tp_{\mu\nu}$, we confirmed that tensor waves propagate at light speed, and carry two independent degrees of freedom which are the standard plus and cross modes of GWs. Scalar waves, however, are generally subluminal, except if $G_{2,XX}=G_{3,X}=0$, i.e., for conformally coupled quintessence. Scalar waves produce curvature perturbations and are thereby measurable in principle; the associated tidal forces are transverse and circularly symmetric for luminal scalar waves, and generally triaxial in the subluminal case.

We found that luminal and quasiluminal scalar waves do not interact with tensor waves. In that case, scalar and tensor waves propagate without seeing each others, just like the horizontal and vertical polarizations of light. Each wave, scalar or tensor, defines its own notion of distance, $D\e{S}$ or $D\e{G}$, which quantifies how its energy dilutes as the waves propagate. The expression of $D\e{G}$, its connection with the electromagnetic luminosity distance, and the resulting discussions about standard sirens, remain unchanged compared to Ref.~\cite{Dalang:2019rke}. Because the emission of scalar radiation is highly constrained by observations, we conclude that luminal and quasiluminal scalar waves are irrelevant to GW physics in reduced Horndeski theories.

The case of frankly subluminal scalar waves is more subtle. We argued that if scalar and tensor waves propagate in the same direction, their interaction is effectively suppressed because their phases are incoherent. However, the general case would deserve a dedicated study which is beyond the scope of the present work. In particular, tensor waves propagating within a subluminal scalar medium should generate scalar shock waves similar to Cherenkov radiation. The analysis of this phenomenon, together with its observational consequences, will be exposed in a subsequent article, for which the present one constitutes a solid basis.

\section{Acknowledgments}
C.D. thanks Fabien Lacasa and Michele Oliosi for inspiring discussions. C.D. and P.F. warmly thank Miguel Zumalacárregui and especially Jose María Ezquiaga for an enlightening discussion which followed the first version of this article, and eventually led to the addition of \cref{subsec:nonluminal}. We also thank Tan Liu, Yan Wang and Wen Zhao for spotting a few typos reported in \cite{Liu:2022qcx}. C.D. and L.L. were supported by a Swiss National Science Foundation (SNSF) Professorship grant (No.~170547). P.F. received the support of a fellowship from ``la Caixa'' Foundation (ID 100010434). The fellowship code is LCF/BQ/PI19/11690018.

\appendix
\begin{widetext}

\section{Explicit expressions involved in the linearized equations of motion}
\label{sec:explicit_EoM}

This appendix provides the long explicit expressions involved in the linearized equations of motion for $\dph, h_{\mu\nu}$. \Cref{subsec:brute_force} gives the result of a brute-force linearization of $\EOM_\ph, \EOM_{\mu\nu}$. \Cref{subsec:diagonal} is the result after diagonalization of the kinetic term and application of the generalized harmonic gauge~\eqref{eq:harmonic_gauge}.

The formulas provided in this appendix have been independently derived and cross-checked by the first two authors of this article, so as to mitigate the high risk of computational errors and typos.

\subsection{Before diagonalization}
\label{subsec:brute_force}

Recall that we have classified the various terms arising from the equations of motion into three categories: (i) kinetic terms, with second derivatives of $\dph, h_{\mu\nu}$; (ii) amplitude terms, with first derivatives; and (iii) masslike terms, with no derivatives. In practice, this classification takes the following form:
\begin{equation}
\label{eq:undiagonalized_EoM_appendix}
\begin{pmatrix}
\delta\EOM_\ph\\
\delta\EOM_{\mu\nu}
\end{pmatrix}
=
\Bigg[
    \underbrace{
    \begin{pmatrix}
    K\indices{_\ph^\ph^\alpha^\beta} &
    K\indices{_\ph^\rho^\sigma^\alpha^\beta} \\
    K\indices{_\mu_\nu^\ph^\alpha^\beta} &
    K\indices{_\mu_\nu^\rho^\sigma^\alpha^\beta}
    \end{pmatrix}
    }_{\text{kinetic matrix }\kineticmatrix^{\alpha\beta}}
    \bar{\nabla}_\alpha
    \bar{\nabla}_\beta
    +
    \underbrace{
    \begin{pmatrix}
    A\indices{_\ph^\ph^\alpha} &
    A\indices{_\ph^\rho^\sigma^\alpha} \\
    A\indices{_\mu_\nu^\ph^\alpha} &
    A\indices{_\mu_\nu^\rho^\sigma^\alpha}
    \end{pmatrix}
    }_{\text{amplitude matrix }\amplitudematrix^\alpha}
    \bar{\nabla}_\alpha
    +
    \underbrace{
    \begin{pmatrix}
    M\indices{_\ph^\ph} &
    M\indices{_\ph^\rho^\sigma} \\
    M\indices{_\mu_\nu^\ph} &
    M\indices{_\mu_\nu^\rho^\sigma}
    \end{pmatrix}
    }_{\text{mass matrix }\massmatrix}
\Bigg]
\begin{pmatrix}
\dph\\
\hh_{\rho\sigma}
\end{pmatrix} .
\end{equation}
We give hereafter the explicit expressions of the kinetic and amplitude blocks, but not of the masslike terms which are neglected. Each block generically receives contributions from the three Lagrangians $\Lagrangian_2, \Lagrangian_3, \Lagrangian_4$; hence, we shall write for instance
\begin{equation}
K\indices{_\ph^\ph^\alpha^\beta}
= \sum_{i=2}^4 \tensor[^{(i)}]{K}{_\ph^\ph^\alpha^\beta} \ ,
\end{equation}
where a superscript $(i)$ indicates that the associated term comes from $\Lagrangian_i$; the same terminology applies to the other blocks.

When giving the expressions of the matrix blocks, we may choose to contract or not their indices, depending on what provides the best readability.

\subsubsection{Blocks of the kinetic matrix}
\label{subsubsec:kinetic_matrix_nondiagonal}

\begin{description}
\item[Scalar kinetic terms in the scalar equation of motion]
$K\indices{_\ph^\ph^\alpha^\beta}\dph_{;\alpha\beta}$
\begin{align}
\tensor[^{(2)}]{K}{_\ph^\ph^\alpha^\beta}
&=
\bG_{2,X} \bg^{\alpha\beta}
- \bG_{2,XX} \bph^{,\alpha} \bph^{,\beta}
\\
\tensor[^{(3)}]{K}{_\ph^\ph^\alpha^\beta}
&=
\pa{
    2 \bG_{3,\ph}
    - 2 \bar{X} \bG_{3,\ph X}
    +  \bG_{3,XX} \bph^{,\mu}\bar{X}_{,\mu}
    + 2 \bG_{3,X} \Box \bph
    } \bg^{\alpha\beta} \nonumber\\
&\quad
- \pa{
    \bG_{3,XX} \bph^{,\alpha}\bph^{,\beta}\Box \bph
    + 2 \bG_{3,\ph X} \bph^{,\alpha} \bph^{,\beta}
    + 2 \bG_{3,XX} \bar{X}^{,(\alpha} \bph^{,\beta)}
    + 2 \bG_{3,X} \bph^{;\alpha\beta}
    }
\\
\tensor[^{(4)}]{K}{_\ph^\ph^\alpha^\beta}
&= 0
\end{align}

\item[Metric kinetic terms in the scalar equation of motion] $K\indices{_\ph^{\rho\sigma}^\alpha^\beta} \hh_{\rho\sigma;\alpha\beta}$
\begin{align}
\tensor[^{(2)}]{K}{_\ph^{\rho\sigma}^\alpha^\beta}
\hh_{\rho\sigma;\alpha\beta}
&= 0
\\
\tensor[^{(3)}]{K}{_\ph^{\rho\sigma}^\alpha^\beta}
\hh_{\rho\sigma;\alpha\beta}
&= - \bG_{3,X} \bph^{,\rho}\bph^{,\sigma}
    \delta R_{\rho\sigma}
= - \frac{1}{2} \bG_{3,X}
    \pac{
        2\bg^{\rho\alpha}\bph^{,\sigma}\bph^{,\beta}
        - \pa{
            \bph^{,\rho}\bph^{,\sigma}
            + \bX \bg^{\rho\sigma}
            } \bg^{\alpha\beta}
        }
    \hh_{\rho\sigma;\alpha\beta}
\\
\tensor[^{(4)}]{K}{_\ph^{\rho\sigma}^\alpha^\beta}
\hh_{\rho\sigma;\alpha\beta}
&= \bG_{4,\ph} \delta R
= \frac{1}{2} \bG_{4,\ph}
    \pa{
        2\bg^{\rho\alpha}\bg^{\sigma\beta}
        + \bg^{\rho\sigma} \bg^{\alpha\beta}
        } \hh_{\rho\sigma;\alpha\beta}
\end{align}

\item[Scalar kinetic terms in the metric equation of motion] $K\indices{_{\mu\nu}^\ph^\alpha^\beta}\dph_{;\alpha\beta}$
\begin{align}
\tensor[^{(2)}]{K}{_{\mu\nu}^\ph^\alpha^\beta}
&= 0
\\
\tensor[^{(3)}]{K}{_{\mu\nu}^\ph^\alpha^\beta}
&= \frac{1}{2} \bG_{3,X}
    \pac{
        2\bph_{,(\mu}\bph^{,\alpha} \delta^\beta_{\nu)}
        - \bph_{,\mu} \bph_{,\nu} \bg^{\alpha\beta}
        - \bg_{\mu\nu} \bph^{,\alpha} \bph^{,\beta}
        }
\\
\tensor[^{(4)}]{K}{_{\mu\nu}^\ph^\alpha^\beta}
&= \bG_{4,\ph}
    \pac{
        \bg_{\mu\nu}\bg^{\alpha\beta}
        - \delta^{(\alpha}_\mu \delta^{\beta)}_\nu
        }
\end{align}

\item[Metric kinetic terms in the metric equation of motion] $K\indices{_{\mu\nu}^{\rho\sigma}^\alpha^\beta}\hh_{\rho\sigma;\alpha\beta}$
\begin{align}
\tensor[^{(2)}]{K}{_{\mu\nu}^{\rho\sigma}^\alpha^\beta}
\hh_{\rho\sigma;\alpha\beta}
&= 0
\\
\tensor[^{(3)}]{K}{_{\mu\nu}^{\rho\sigma}^\alpha^\beta}
\hh_{\rho\sigma;\alpha\beta}
&= 0
\\
\tensor[^{(4)}]{K}{_{\mu\nu}^{\rho\sigma}^\alpha^\beta}
\hh_{\rho\sigma;\alpha\beta}
&= G_4 \, \delta E_{\mu\nu}
= \frac{G_4}{2}
    \pac{
        2\hh\indices{_\rho_(_\mu^;^\rho_\nu_)}
        - \Box\hh_{\mu\nu}
        - \hh\indices{^\rho^\sigma_;_\rho_\sigma}
            \bg_{\mu\nu}
        }
\end{align}
\end{description}

\subsubsection{Blocks of the amplitude matrix}
\label{subsubsec:amplitude_matrix_nondiagonal}

\begin{description}
\item[Scalar amplitude terms in the scalar equation of motion] $A\indices{_\ph^\ph^\alpha}\dph_{,\alpha}$
\begin{align}
\tensor[^{(2)}]{A}{_\ph^\ph^\alpha}
&= \pa{
        - G_{2,XX} \Box\bph
        + G_{2,X\ph}
        + 2\bX G_{2,XX\ph}
        - G_{2,XXX} \bph^{,\mu}\bX_{,\mu}
        } \bph^{,\alpha}
    + 2 G_{2,XX} \bX^{,\alpha}
\\
\tensor[^{(3)}]{A}{_\ph^\ph^\alpha}
&=
\Big[
    \pa{
        2\bar{X} \bG_{3,\ph X X} - \bG_{3,XXX} \bph_{,\mu} \bar{X}^{,\mu} - \bG_{3,XX} \Box\bph
        } \Box\bph
    + 2\bG_{3,\ph\ph}
    + 2\bar{X}\bG_{3,\ph\ph X}
    - 2\bG_{3,\ph XX}\bph_{,\mu} \bar{X}^{,\mu}
    - \bG_{3,XXX} \bar{X}_{,\mu} \bar{X}^{,\mu}
\nonumber\\&\quad
    + \bG_{3,XX}
        \pa{
            \bph^{;\mu\nu} \bph_{;\mu\nu} + \bar{R}^{\mu\nu}\bph_{,\mu}\bph_{,\nu}
            }
\Big] \bph^{,\alpha}
+
\Big[
    2\bG_{3,XX} \Box\bph + 4\bG_{3,\ph X}
\Big] \bar{X}^{,\alpha}
- 2 \bG_{3,XX} \bar{X}_{,\mu} \bph^{;\mu\alpha}
- 2 \bG_{3,X} \bar{R}^{\mu\alpha} \bph_{,\mu}
\\
\tensor[^{(4)}]{A}{_\ph^\ph^\alpha}
&= 0
\end{align}

\item[Metric amplitude terms in the scalar equation of motion] $A\indices{_\ph^{\rho\sigma}^\alpha}\hh_{\rho\sigma;\alpha}$
\begin{align}
\tensor[^{(2)}]{A}{_\ph^{\rho\sigma}^\alpha}
&= - \bG_{2,X} \bph^{,\rho}\bg^{\sigma\alpha}
     + \frac{1}{2} \bG_{2,XX}
        \pa{
            \bph^{,\rho}\bph^{,\sigma}
            + \bX\bg^{\rho\sigma}
            } \bph^{,\alpha}
\\
\tensor[^{(3)}]{A}{_\ph^{\rho\sigma}^\alpha}
&= -\pac{
        2\bG_{3,\ph}
        - 2\bX \bG_{3,\ph X}
        + \bG_{3,XX} \bph^{,\mu}\bX_{,\mu}
        + 2\bG_{3,X} \Box\bph
        } \bph^{,\rho}\bg^{\sigma\alpha}
\nonumber\\&\quad
    + \pac{
        \pa{ \bG_{3,\ph X} + \frac{1}{2}\bG_{3,XX}\Box\bph }
            \bph^{,\sigma}
        + \bG_{3,XX} \bX^{,\alpha}
        } \pa{\bph^{,\rho}\bph^{,\sigma}+\bX\bg^{\rho\sigma}}
\nonumber\\&\quad    
        + \bG_{3,X}
            \pac{
                2\bph^{;\alpha(\rho}\bph^{,\sigma)}
                - \bph^{;\alpha\lambda}\bph_{,\lambda}\bg^{\rho\sigma}
                - \bph^{;\rho\sigma}\bph^{,\alpha}
                + \frac{1}{2}\Box\bph\bph^{,\alpha}\bg^{\rho\sigma}
                }
\\
\tensor[^{(4)}]{A}{_\ph^{\rho\sigma}^\alpha}
&= 0
\end{align}

\item[Scalar amplitude terms in the metric equation of motion] $A\indices{_{\mu\nu}^\ph^\alpha}\dph_{,\alpha}$
\begin{align}
\tensor[^{(2)}]{A}{_{\mu\nu}^\ph^\alpha}
&= \frac{1}{2}
    \pa{
        \bG_{2,XX}\bph_{,\mu}\bph_{,\nu}
        + \bG_{2,X} \bg_{\mu\nu}
        } \bph^{,\alpha}
    - \bG_{2,X} \bph_{,(\mu} \delta^\alpha_{\nu)}
\\
\tensor[^{(3)}]{A}{_{\mu\nu}^\ph^\alpha}
&= \bar{g}_{\mu\nu}
    \pac{
        \pa{
            \bG_{3,\ph}
            + \bar{X} \bG_{3,\ph X}
            - \frac{1}{2} \bG_{3,XX}
                \bph^{,\rho} \bar{X}_{,\rho}
            } \bph^{,\alpha}
        + \bG_{3,X} \bar{X}^{,\alpha}
        }
    + \bph_{,\mu} \bph_{,\nu}
        \pa{
        \bG_{3,\ph X}
        + \frac{1}{2}\bG_{3,XX}\Box \bph
        } \bph^{,\alpha}
    \nonumber\\&\quad
    - \pac{
            2\bG_{3,\ph} \bph_{,(\mu }
            + \bG_{3,X} \bar{X}_{,(\mu}
            + \bG_{3,X} \Box \bph \bph_{,(\mu }
            } \delta_{,\nu)}^\alpha
    + \bG_{3,XX}\bph_{,(\mu} \bar{X}_{,\nu)}
        \bph^{,\alpha}
    + \bG_{3,X} \bph_{,(\mu}\tensor{\bph}{_;_\nu_)^\alpha}
\\
\tensor[^{(4)}]{A}{_{\mu\nu}^\ph^\alpha}
&= 2\bG_{4,\ph\ph}
    \pac{
        \bg_{\mu\nu} \bph^{,\alpha}
        - \bph_{(\mu}\delta^\alpha_{\nu)}
        }
\end{align}

\item[Metric amplitude terms in the metric equation of motion] $A\indices{_{\mu\nu}^{\rho\sigma}^\alpha}\hh_{\rho\sigma;\alpha}$
\begin{align}
\tensor[^{(2)}]{A}{_{\mu\nu}^{\rho\sigma}^\alpha}
&= 0
\\
\tensor[^{(3)}]{A}{_{\mu\nu}^{\rho\sigma}^\alpha}
&= \frac{1}{2} \bG_{3,X}
    \pac{
        \pa{
            \bph^{,\rho}\bph^{,\sigma}
            + \bX\bg^{\rho\sigma}
            }
        \pa{
            \frac{1}{2}\bg_{\mu\nu}\bph^{,\alpha}
            - \bph_{,(\mu}\delta^{\alpha}_{,\nu)}
            }
        + \bph_{,\mu}\bph_{,\nu}
            \bph^{,(\rho}\bg^{\sigma)\alpha}
        }
\\
\tensor[^{(4)}]{A}{_{\mu\nu}^{\rho\sigma}^\alpha}
&= \frac{1}{2} \bG_{4,\ph}
    \pac{
        \pa{
            2\bph^{,(\rho}\delta^{\sigma)}_{(\mu}\delta^\alpha_{\nu)}
            - \bph_{,(\mu}\delta^\alpha_{,\nu)}\bg^{\rho\sigma}
            }
        - \pa{
            \delta^\rho_{(\mu}\delta^\sigma_{\nu)}
            - \frac{1}{2}\bg_{\mu\nu}\bg^{\rho\sigma}
            } \bph^{,\alpha}
        - 2\bg_{\mu\nu} \bph^{,(\rho}\bg^{\sigma)\alpha}
        }
\end{align}
The last two matrix elements being hard to read, we also provide their contracted counterparts:
\begin{align}
\tensor[^{(3)}]{A}{_{\mu\nu}^{\rho\sigma}^\alpha}
\hh_{\rho\sigma;\alpha}
&= \frac{1}{2} \bG_{3,X}
    \pac{
        \pa{
            \bph^{,\rho}\bph^{,\sigma}
            + \bX\bg^{\rho\sigma}
            }
        \pa{
            \frac{1}{2}\bg_{\mu\nu}\bph^{,\alpha}
                \hh_{\rho\sigma;\alpha}
            - \hh_{\rho\sigma;(\nu} \bph_{,\mu)}
            }
        + \bph_{,\mu} \bph_{,\nu} \bph_{,\rho}
            \hh\indices{^\rho^\sigma_;_\sigma}
        }
\\
\tensor[^{(4)}]{A}{_{\mu\nu}^{\rho\sigma}^\alpha}
\hh_{\rho\sigma;\alpha}
&= \frac{1}{2} \bG_{4,\ph} \bph^{,\rho}
    \pa{
        2\hh_{\rho(\mu;\nu)}
        - \bg_{\rho(\mu} \hh_{,\nu)}
        - \hh_{\mu\nu;\rho}
        + \frac{1}{2} \bg_{\mu\nu} \hh_{,\rho}
        - 2\bg_{\mu\nu} \hh\indices{_\rho_\sigma^;^\sigma}
        } .
\end{align}
\end{description}

\subsection{After diagonalization and harmonic-gauge fixing}
\label{subsec:diagonal}

\subsubsection{Description of the operations}
\label{subsubsec:diagonalization_operations}

The kinetic term of the differential system~\eqref{eq:undiagonalized_EoM_appendix} can be diagonalized and simplified by applying the following operations:
\begin{enumerate}
\item\textit{Elimination of $K\indices{_\ph^\rho^\sigma^\alpha^\beta}$.} In this first step, we substitute the Ricci-curvature terms that appear in the scalar equation of motion~$\delta\EOM_{\ph}$ using the tensor equation of motion~$\delta\EOM_{\mu\nu}$. This operation is possible because $K\indices{_\mu_\nu^\rho^\sigma^\alpha^\beta}h_{\mu\nu;\alpha\beta}\propto\delta E_{\mu\nu}=\delta\hat{R}_{\mu\nu}$. In other words, the terms containing $h_{\rho\sigma;\alpha\beta}$ in the scalar equation of motion are linear combinations of the terms $h_{\rho\sigma;\alpha\beta}$ in the tensor equation of motion. To be really specific, we have
\begin{align}
K\indices{_\ph^\rho^\sigma^\alpha^\beta}\hh_{\rho\sigma;\alpha\beta}
&= \tensor[^{(3)}]{K}{_\ph^\rho^\sigma^\alpha^\beta}\hh_{\rho\sigma;\alpha\beta}
    + \tensor[^{(4)}]{K}{_\ph^\rho^\sigma^\alpha^\beta}\hh_{\rho\sigma;\alpha\beta}
\\
&= -\bG_{3,X} \bph^{,\mu}\bph^{,\nu} \delta R_{\mu\nu}
    + \bG_{4,\ph} \delta R
\\
&= - \pac{
        \bG_{3,X} \pa{
                    \bph^{,\mu}\bph^{,\nu}
                    + X\bg^{\mu\nu}
                    }
        + \bG_{4,\ph} \bg^{\mu\nu}
        }
    \delta E_{\mu\nu}
\\
&= - \pac{
        \bG_{3,X} \pa{
                    \bph^{,\mu}\bph^{,\nu}
                    + X\bg^{\mu\nu}
                    }
        + \bG_{4,\ph} \bg^{\mu\nu}
        }
    \pa{
    \bG_4^{-1}
    K\indices{_\mu_\nu^\rho^\sigma^\alpha^\beta}\hh_{\rho\sigma;\alpha\beta}
    }
\\
&= \underbrace{
    \bG_{4}^{-1}
    \pac{
        \bG_{3,X} \pa{
                    \bph^{,\mu}\bph^{,\nu}
                    + X\bg^{\mu\nu}
                    }
        + \bG_{4,\ph} \bg^{\mu\nu}
        }
    }_{\define C^{\mu\nu}}
    \pa{
        K\indices{_\mu_\nu^\ph^\alpha^\beta}\dph_{;\alpha\beta}
        + A\indices{_\mu_\nu^\ph^\alpha}\dph_{,\alpha}
        + A\indices{_\mu_\nu^\rho^\sigma^\alpha}\hh_{\rho\sigma;\alpha}
        }
\end{align}

The tensor equation of motion, $\delta\EOM_{\mu\nu}=0$, was used to go from the penultimate to the last line. When the above formula is substituted into the scalar equation of motion, the right-hand side contributes to the diagonal kinetic term and to the amplification matrix. Precisely, the following transformations apply:
\begin{align}
K\indices{_\ph^\rho^\sigma^\alpha^\beta}
&\longmapsto 0
\\
K\indices{_\ph^\ph^\alpha^\beta}
&\longmapsto K\indices{_\ph^\ph^\alpha^\beta}
    + C^{\mu\nu} K\indices{_\mu_\nu^\ph^\alpha^\beta}
\\
A\indices{_\ph^\ph^\alpha}
&\longmapsto A\indices{_\ph^\ph^\alpha}
    + C^{\mu\nu}A\indices{_\mu_\nu^\ph^\alpha}
\\
A\indices{_\ph^\rho^\sigma^\alpha}
&\longmapsto A\indices{_\ph^\rho^\sigma^\alpha}
    + C^{\mu\nu} A\indices{_\mu_\nu^\rho^\sigma^\alpha} \ ,
\end{align}
or, in matrix terms,
\begin{align}
\kineticmatrix^{\alpha\beta}
\longmapsto
\begin{pmatrix}
1 & C^{\mu\nu}\\
0 & 1
\end{pmatrix}
\kineticmatrix^{\alpha\beta}
\qquad\text{and}\qquad
\amplitudematrix^{\alpha}
\longmapsto
\begin{pmatrix}
1 & C^{\mu\nu}\\
0 & 1
\end{pmatrix}
\amplitudematrix^{\alpha}
\end{align}
which is an operation on the rows of the matrices.

We note that the additional terms to $K\indices{_\ph^\ph^\alpha^\beta}, A\indices{_\ph^\ph^\alpha}, A\indices{_\ph^\rho^\sigma^\alpha}$ are all quadratic in the coupling functions, i.e.~they feature pre-factors such as $G_{3,X}G_{4,\ph\ph}$, or $G_{3,X}^2$, etc. Because of that, they can no longer be associated with a specific Lagrangian~$\Lagrangian_2, \Lagrangian_3, \Lagrangian_4$, and hence do not fit into the classification that we have used to far. All these hybrid terms will therefore be gathered under the label (5) in the final result.

\item\textit{Elimination of $K\indices{_\mu_\nu^\ph^\alpha^\beta}$.} While the previous step consisted in combining the equations of motion, i.e.~acting on the rows of the system matrices, this second step will consist in mixing its variables, i.e.~act on the columns of the system matrices. Namely, we introduce the new tensor variable
\begin{equation}
\tp_{\mu\nu} \define \hh_{\mu\nu} + \hat{C}_{\mu\nu} \delta\ph \ ,
\qquad
\hat{C}_{\mu\nu} = \bG_{4}^{-1}\pa{ \bG_{3,X}\bph_{,\mu}\bph_{,\nu} -\bG_{4,\ph} \bg_{\mu\nu} }
\end{equation}
being the trace-reversed counterpart of the tensor $C_{\mu\nu}$ that appeared in the previous step. Substituting, in the tensor equation of motion, $h_{\rho\sigma;\alpha\beta}$ by its expression in terms of $\tp_{\mu\nu}, \dph$, eliminates the off-diagonal kinetic term $K\indices{_\mu_\nu^\ph^\alpha^\beta}$. This cancellation is due to the remarkable identity
$
K\indices{_\mu_\nu^\rho^\sigma^\alpha^\beta} \hat{C}_{\rho\sigma}
= K\indices{_\mu_\nu^\ph^\alpha^\beta}
$,
so that, up to masslike terms,
\begin{align}
K\indices{_\mu_\nu^\rho^\sigma^\alpha^\beta}\hh_{\rho\sigma;\alpha\beta}
+ K\indices{_\mu_\nu^\ph^\alpha^\beta}\dph_{;\alpha\beta}
&= K\indices{_\mu_\nu^\rho^\sigma^\alpha^\beta}
    \pa{
        \tp_{\rho\sigma;\alpha\beta}
        - \hat{C}_{\rho\sigma}\dph_{;\alpha\beta}
        - 2\hat{C}_{\rho\sigma;(\alpha}\dph_{,\beta)}
        }
+ K\indices{_\mu_\nu^\ph^\alpha^\beta}\dph_{;\alpha\beta}
\\
&= K\indices{_\mu_\nu^\rho^\sigma^\alpha^\beta} \tp_{\rho\sigma;\alpha\beta}
    -2K\indices{_\mu_\nu^\rho^\sigma^\alpha^\beta}\hat{C}_{\rho\sigma;(\alpha}\dph_{,\beta)} \ .
\end{align}

The change from $\hh_{\mu\nu}$ to $\tp_{\mu\nu}$ also directly changes the amplitude matrix, since (again up to masslike terms)
\begin{align}
A\indices{_\mu_\nu^\rho^\sigma^\alpha}
\hh_{\rho\sigma;\alpha}
&= A\indices{_\mu_\nu^\rho^\sigma^\alpha}
    \pa{
        \tp_{\rho\sigma;\alpha}
        - \hat{C}_{\rho\sigma}\dph_{,\alpha}
        } ,
\\
A\indices{_\ph^\rho^\sigma^\alpha}
\hh_{\rho\sigma;\alpha}
&= A\indices{_\mu_\nu^\ph^\alpha}
    \pa{
        \tp_{\rho\sigma;\alpha}
        - \hat{C}_{\rho\sigma}\dph_{,\alpha}
        } .
\end{align}
Summarizing, this second operation leads to the following transformations:
\begin{align}
K\indices{_\mu_\nu^\ph^\alpha^\beta}
&\longmapsto 0\,,
\\
\label{eq:transformation_A_h^phi_step_2}
A\indices{_\mu_\nu^\ph^\alpha}
&\longmapsto
A\indices{_\mu_\nu^\ph^\alpha}
- 2K\indices{_\mu_\nu^\rho^\sigma^(^\alpha^\beta^)} \hat{C}_{\rho\sigma;\beta}
- A\indices{_\mu_\nu^\rho^\sigma^\alpha} \hat{C}_{\rho\sigma}\,,
\\
\label{eq:transformation_A_phi^phi_step_2}
A\indices{_\ph^\ph^\alpha}
&\longmapsto
A\indices{_\ph^\ph^\alpha}
- A\indices{_\ph^\rho^\sigma^\alpha} \hat{C}_{\rho\sigma}\,,
\end{align}
where we must not forget to account for the modification of $A\indices{_\ph^\rho^\sigma^\alpha}$ that occurred in step 1. In matrix terms, that second operation reads
\begin{equation}
\kineticmatrix^{\alpha\beta}
\longmapsto
\kineticmatrix^{\alpha\beta}
\begin{pmatrix}
1 & 0\\
-\hat{C}_{\rho\sigma} & 1
\end{pmatrix}
\qquad\text{and}\qquad
\amplitudematrix^{\alpha}
\longmapsto
\pac{
\amplitudematrix^{\alpha}
+
2 \kineticmatrix^{(\alpha\beta)}
\bar{\nabla}_\beta
}
\begin{pmatrix}
1 & 0\\
-\hat{C}_{\rho\sigma} & 1
\end{pmatrix} .
\end{equation}

In \cref{eq:transformation_A_h^phi_step_2}, the correction coming from the kinetic matrix is linear in the coupling functions, and hence it directly changes $\tensor[^{(3)}]{A}{_\mu_\nu^\ph^\alpha}, \tensor[^{(4)}]{A}{_\mu_\nu^\ph^\alpha}$. All the other corrections are quadratic in the coupling functions, and thus will be classified under the label (5) as the modifications originating from the step 1.

\item\textit{Imposing the harmonic gauge.} Impose the harmonic gauge $\tp\indices{^\mu^\nu_;_\nu}=0$, which removes a few terms in the equations of motion. In particular, this drastically simplifies the kinetic term for $\tp_{\mu\nu}$.
\end{enumerate}

The resulting system then reads
\begin{equation}
\Bigg[
    \underbrace{
    \begin{pmatrix}
    \diag{K}\indices{_\ph^\ph^\alpha^\beta} &
    0 \\
    0 &
    \diag{K}\indices{_\mu_\nu^\rho^\sigma^\alpha^\beta}
    \end{pmatrix}
    }_{\text{diagonal kinetic matrix }\diag{\kineticmatrix}^{\alpha\beta}}
    \bar{\nabla}_\alpha
    \bar{\nabla}_\beta
    +
    \underbrace{
    \begin{pmatrix}
    \diag{A}\indices{_\ph^\ph^\alpha} &
    \diag{A}\indices{_\ph^\rho^\sigma^\alpha} \\
    \diag{A}\indices{_\mu_\nu^\ph^\alpha} &
    \diag{A}\indices{_\mu_\nu^\rho^\sigma^\alpha}
    \end{pmatrix}
    }_{\text{new amplitude matrix }\diag{\amplitudematrix}^\alpha}
    \bar{\nabla}_\alpha
\Bigg]
\begin{pmatrix}
\dph\\
\tp_{\rho\sigma}
\end{pmatrix}
=
\begin{pmatrix}
0\\
0
\end{pmatrix} ,
\end{equation}
and the various terms generically decompose as, e.g.,
\begin{equation}
K\indices{_\ph^\ph^\alpha^\beta}
= \sum_{i=2}^5 \tensor[^{(i)}]{K}{_\ph^\ph^\alpha^\beta} \ ,
\end{equation}
where the terms with the label (5) are at least quadratic in the coupling functions.

\subsubsection{Blocks of the kinetic matrix}
\label{subsubsec:kinetic_matrix_diagonal}

\begin{description}
\item[Scalar kinetic terms in the scalar equation of motion]
$\diag{K}\indices{_\ph^\ph^\alpha^\beta}\dph_{;\alpha\beta}$
\begin{align}
\tensor[^{(2)}]{\diag{K}}{_\ph^\ph^\alpha^\beta}
&= \tensor[^{(2)}]{K}{_\ph^\ph^\alpha^\beta}
=
\bG_{2,X} \bg^{\alpha\beta}
- \bG_{2,XX} \bph^{,\alpha} \bph^{,\beta}
\\
\tensor[^{(3)}]{\diag{K}}{_\ph^\ph^\alpha^\beta}
&= \tensor[^{(3)}]{K}{_\ph^\ph^\alpha^\beta}
=
\pa{
    2 \bG_{3,\ph}
    - 2 \bar{X} \bG_{3,\ph X}
    +  \bG_{3,XX} \bph^{,\mu}\bar{X}_{,\mu}
    + 2 \bG_{3,X} \Box \bph
    } \bg^{\alpha\beta} \nonumber\\
&\hspace{2.5cm}
- \pa{
    \bG_{3,XX} \bph^{,\alpha}\bph^{,\beta}\Box \bph
    + 2 \bG_{3,\ph X} \bph^{,\alpha} \bph^{,\beta}
    + 2 \bG_{3,XX} \bar{X}^{,\alpha} \bph^{,\beta}
    + 2 \bG_{3,X} \bph^{;\alpha\beta}
    }
\\
\tensor[^{(4)}]{\diag{K}}{_\ph^\ph^\alpha^\beta}
&= \tensor[^{(4)}]{K}{_\ph^\ph^\alpha^\beta}
= 0
\\
\tensor[^{(5)}]{\diag{K}}{_\ph^\ph^\alpha^\beta}
&= C^{\mu\nu} K\indices{_\mu_\nu^\ph^\alpha^\beta}
=
\bG_{4}^{-1}
    \pac{
        \pa{
            3\bG_{4,\ph}^2
            + 2\bX\bG_{3,X}\bG_{4,\ph}
            - \bX^2\bG_{3,X}^2
            } \bg^{\alpha\beta}
        - 2\bG_{3,X} \pa{\bG_{4,\ph}+\bX\bG_{3,X}}
            \bph^{,(\alpha} \bph^{,\beta)}
        }
\end{align}

\item[Metric kinetic terms in the metric equation of motion] $\diag{K}\indices{_{\mu\nu}^{\rho\sigma}^\alpha^\beta}\tp_{\rho\sigma;\alpha\beta}$
\begin{align}
\tensor[^{(2)}]{\diag{K}}{_{\mu\nu}^{\rho\sigma}^\alpha^\beta}
&= \tensor[^{(2)}]{K}{_{\mu\nu}^{\rho\sigma}^\alpha^\beta}
= 0
\\
\tensor[^{(3)}]{\diag{K}}{_{\mu\nu}^{\rho\sigma}^\alpha^\beta}
&= \tensor[^{(3)}]{K}{_{\mu\nu}^{\rho\sigma}^\alpha^\beta}
= 0
\\
\tensor[^{(4)}]{\diag{K}}{_{\mu\nu}^{\rho\sigma}^\alpha^\beta}
&= -\frac{1}{2} G_4
        \delta^\rho_{,(\mu}
        \delta^\sigma_{,\nu)}
        \bg^{\alpha\beta}
\qquad \text{i.e.} \qquad
\tensor[^{(4)}]{\diag{K}}{_{\mu\nu}^{\rho\sigma}^\alpha^\beta}
\tp_{\rho\sigma;\alpha\beta}
= -\frac{1}{2} G_4 \Box \tp_{\mu\nu}
\end{align}
\end{description}

\subsubsection{Blocks of the amplitude matrix}
\label{subsubsec:amplitude_matrix_diagonal}

\begin{description}
\item[Scalar amplitude terms in the scalar equation of motion] $\diag{A}\indices{_\ph^\ph^\alpha}\dph_{,\alpha}$
\begin{align}
\tensor[^{(2)}]{\diag{A}}{_\ph^\ph^\alpha}
&= \tensor[^{(2)}]{A}{_\ph^\ph^\alpha}
= \pa{
        - G_{2,XX} \Box\bph
        + G_{2,X\ph}
        + 2\bX G_{2,XX\ph}
        - G_{2,XXX} \bph^{,\mu}\bX_{,\mu}
        } \bph^{,\alpha}
    + 2 G_{2,XX} \bX^{,\alpha}
\\
\tensor[^{(3)}]{\diag{A}}{_\ph^\ph^\alpha}
&= \tensor[^{(3)}]{A}{_\ph^\ph^\alpha} \nonumber\\
&=
\Big[
    \pa{
        2\bar{X} \bG_{3,\ph X X} - \bG_{3,XXX} \bph_{,\mu} \bar{X}^{,\mu} - \bG_{3,XX} \Box\bph
        } \Box\bph
    + 2\bG_{3,\ph\ph}
    + 2\bar{X}\bG_{3,\ph\ph X}
    - 2\bG_{3,\ph XX}\bph_{,\mu} \bar{X}^{,\mu}
    - \bG_{3,XXX} \bar{X}_{,\mu} \bar{X}^{,\mu}
\nonumber\\&\quad
    + \bG_{3,XX}
        \pa{
            \bph^{;\mu\nu} \bph_{;\mu\nu} + \bar{R}^{\mu\nu}\bph_{,\mu}\bph_{,\nu}
            }
\Big] \bph^{,\alpha}
+
\Big[
    2\bG_{3,XX} \Box\bph + 4\bG_{3,\ph X}
\Big] \bar{X}^{,\alpha}
- 2 \bG_{3,XX} \bar{X}_{,\mu} \bph^{;\mu\alpha}
- 2 \bG_{3,X} \bar{R}^{\mu\alpha} \bph_{,\mu}
\\
\tensor[^{(4)}]{\diag{A}}{_\ph^\ph^\alpha}
&= \tensor[^{(4)}]{A}{_\ph^\ph^\alpha}
= 0
\\
\tensor[^{(5)}]{\diag{A}}{_\ph^\ph^\alpha}
&= C^{\mu\nu} A\indices{_\mu_\nu^\ph^\alpha}
    - A\indices{_\ph^\rho^\sigma ^\alpha} \hat{C}_{\rho\sigma}
    - C^{\mu\nu} A\indices{_\mu_\nu^\rho^\sigma^\alpha} \hat{C}_{\rho\sigma}
\end{align}

\item[Metric amplitude terms in the scalar equation of motion] $\diag{A}\indices{_\ph^{\rho\sigma}^\alpha}\tp_{\rho\sigma;\alpha}$. The terms (2)-(4) slightly simplify because of the harmonic gauge, which removes any contraction of $\rho,\sigma$ with $\alpha$.
\begin{align}
\tensor[^{(2)}]{\diag{A}}{_\ph^{\rho\sigma}^\alpha}
&= \frac{1}{2} \bG_{2,XX}
    \pa{
        \bph^{,\rho}\bph^{,\sigma}
        + \bX\bg^{\rho\sigma}
        } \bph^{,\alpha}
\\
\tensor[^{(3)}]{\diag{A}}{_\ph^{\rho\sigma}^\alpha}
&= \pac{
        \pa{ \bG_{3,\ph X} + \frac{1}{2}\bG_{3,XX}\Box\bph }
        \bph^{,\alpha}
        + \bG_{3,XX} \bX^{,\alpha}
        } \pa{\bph^{,\rho}\bph^{,\sigma}+\bX\bg^{\rho\sigma}}
\nonumber\\&\quad    
        + \bG_{3,X}
            \pa{
                2\bph^{;\alpha(\rho}\bph^{,\sigma)}
                - \bph^{;\alpha\lambda}\bph_{,\lambda}\bg^{\rho\sigma}
                - \bph^{;\rho\sigma}\bph^{,\alpha}
                + \frac{1}{2}\Box\bph\bph^{,\alpha}\bg^{\rho\sigma}
                }
\\
\tensor[^{(4)}]{\diag{A}}{_\ph^{\rho\sigma}^\alpha}
&= 0
\\
\tensor[^{(5)}]{\diag{A}}{_\ph^{\rho\sigma}^\alpha}
&= C^{\mu\nu} A\indices{_\mu_\nu^\rho^\sigma^\alpha} 
\end{align}

\item[Scalar amplitude terms in the metric equation of motion] $\diag{A}\indices{_{\mu\nu}^\ph^\alpha}\dph_{,\alpha}$. The terms (3), (4) change because of the kinetic term that appears in the transformation~\eqref{eq:transformation_A_h^phi_step_2}.
\begin{align}
\tensor[^{(2)}]{\diag{A}}{_{\mu\nu}^\ph^\alpha}
&= \tensor[^{(2)}]{A}{_{\mu\nu}^\ph^\alpha}
= \frac{1}{2}
    \pa{
        \bG_{2,XX}\bph_{,\mu}\bph_{,\nu}
        + \bG_{2,X} \bg_{\mu\nu}
        } \bph^{,\alpha}
    - \bG_{2,X} \bph_{,(\mu} \delta^\alpha_{\nu)}
\\
\tensor[^{(3)}]{\diag{A}}{_{\mu\nu}^\ph^\alpha}
&= \pac{
        \bG_{3,\ph}
        + \bG_{3,X} \Box\bph
        - \bar{X} \bG_{3,\ph X}
        + \frac{1}{2} \bG_{3,XX} \bar{X}^{,\rho}\bph_{,\rho}
        } \bph^{,\alpha} \bar{g}_{\mu\nu}
    + \pac{
         \bG_{3,\ph X}\bph^{,\alpha}
        + \bG_{3,XX} \bar{X}^{,\alpha}
        + \frac{1}{2} \bG_{3,XX}\Box\bph \bph^{,\alpha}
        } \bph_{,\mu} \bph_{,\nu}
    \nonumber\\&\quad
    + \pac{
        - 2\bG_{3,\ph}
        - 2\bG_{3,X}\Box\ph
        + 2\bar{X} \bG_{3,X\ph}
        - \bG_{3,XX} \bar{X}^{,\rho}\bph_{,\rho}
        } \bph_{,(\mu} \delta_{,\nu)}^\alpha
    + 2\bG_{3,X}\bph_{,(\mu}\bph\indices{_;_\nu_)^\alpha}
    - \bG_{3,X}\bph^{,\alpha}\bph_{;\mu\nu}
\\
\tensor[^{(4)}]{\diag{A}}{_{\mu\nu}^\ph^\alpha}
&= \tensor[^{(4)}]{A}{_{\mu\nu}^{\ph^\alpha}} = 0
\\
\tensor[^{(5)}]{\diag{A}}{_{\mu\nu}^\ph^\alpha}
&= - A\indices{_\mu_\nu^\rho^\sigma^\alpha} \hat{C}_{\rho\sigma} - 2 K\indices{_\mu _\nu ^( ^\alpha ^\beta ^)} \hat{C}_{\rho\sigma;\beta} 
\end{align}

\item[Metric amplitude terms in the metric equation of motion] $\diag{A}\indices{_{\mu\nu}^{\rho\sigma}^\alpha}\tp_{\rho\sigma;\alpha}$. The terms (3), (4) slightly simplify because of the harmonic gauge, which removes any contraction of $\rho,\sigma$ with $\alpha$. The diagonalization process does not produce a (5) term.
\begin{align}
\tensor[^{(2)}]{\diag{A}}{_{\mu\nu}^{\rho\sigma}^\alpha}
&= \tensor[^{(2)}]{A}{_{\mu\nu}^{\rho\sigma}^\alpha} = 0
\\
\tensor[^{(3)}]{\diag{A}}{_{\mu\nu}^{\rho\sigma}^\alpha}
&= \frac{1}{2} \bG_{3,X}
    \pa{
        \bph^{,\rho}\bph^{,\sigma}
         + \bX\bg^{\rho\sigma}
        }
    \pa{
        \frac{1}{2}\bg_{\mu\nu}\bph^{,\alpha}
        - \bph_{(\mu}\delta^{\alpha}_{,\nu)}
        }
\\
\tensor[^{(4)}]{\diag{A}}{_{\mu\nu}^{\rho\sigma}^\alpha}
&= \frac{1}{2}\bG_{4,\ph}
    \pac{
        \pa{
            2\bph^{,(\rho}\delta^{\sigma)}_{(\mu}\delta^\alpha_{\nu)}
            - \bph_{,(\mu}\delta^\alpha_{,\nu)}\bg^{\rho\sigma}
            }
        - \pa{
            \delta^\rho_{(\mu}\delta^\sigma_{\nu)}
            - \frac{1}{2}\bg_{\mu\nu}\bg^{\rho\sigma}
            } \bph^{,\alpha}
        }
\\
\tensor[^{(5)}]{\diag{A}}{_{\mu\nu}^{\rho\sigma}^\alpha}
&= 0
\end{align}
Again, since the (3) and (4) terms are hard to read, we provide their contracted counterparts:
\begin{align}
\tensor[^{(3)}]{\diag{A}}{_{\mu\nu}^{\rho\sigma}^\alpha}
\tp_{\rho\sigma;\alpha}
&= \frac{1}{2} \bG_{3,X}
        \pa{
            \bph^{,\rho}\bph^{,\sigma}
            + \bX\bg^{\rho\sigma}
            }
        \pa{
            \frac{1}{2}\bg_{\mu\nu}\bph^{,\alpha}
                \tp_{\rho\sigma;\alpha}
            - \tp_{\rho\sigma;(\nu} \bph_{,\mu)}
            }
\\
\tensor[^{(4)}]{\diag{A}}{_{\mu\nu}^{\rho\sigma}^\alpha}
\tp_{\rho\sigma;\alpha}
&= \frac{1}{2} \bG_{4,\ph} \bph^{,\rho}
    \pa{
        2\tp_{\rho(\mu;\nu)}
        - \bg_{\rho(\mu} \tp_{,\nu)}
        - \tp_{\mu\nu;\rho}
        + \frac{1}{2} \bg_{\mu\nu} \tp_{,\rho}
        } .
\end{align}
\end{description}

\end{widetext}

\bibliography{bibliography}
\end{document}